\documentclass[12pt,preprint]{aastex}
\usepackage{amsmath}
\usepackage{amssymb}
\usepackage{epsfig}

\newcommand{\cms}{\,{\rm cm$^{-2}$}\,}
\newcommand{\cmc}{\,{\rm cm$^{-3}$}\,}
\newcommand{\kms}{\,{\rm km\,s$^{-1}$}\,}
\newcommand{\kmsmpc}{\,{\rm km\,s$^{-1}$\,Mpc$^{-1}$}\,}

\newcommand{\etal}{{ et~al.~}}
\newcommand{\nexpunit}{\,{\rm photons\,s$^{-1}$\,arcsec$^{-2}$}\,}

\newcommand{\ergs}{\,{\rm erg\,s$^{-1}$}\,}
\newcommand{\ergscm}{\,{\rm erg\,s$^{-1}$\,cm$^{-2}$}\,}
\newcommand{\Ms}{M_\odot}
\newcommand{\Zs}{Z_\odot}

\shorttitle{XMM Observation of an X-ray Trail in the Pavo Group}

\begin{document}


\title{XMM-Newton Observation of an X-ray Trail Between the Spiral 
Galaxy NGC~6872 and the Central Elliptical NGC~6876 
in the Pavo Group}

\author{Marie E. Machacek, Paul Nulsen, Liviu Stirbat,
 Christine Jones, and William R. Forman }
\affil{Harvard-Smithsonian Center for Astrophysics, MS67, \\ 
       60 Garden Street, Cambridge, MA 02138 USA 
\email{mmachacek@cfa.harvard.edu, pnulsen@head-cfa.cfa.harvard.edu, 
   lstirbat@fas.harvard.edu, cjf@head-cfa.cfa.harvard.edu, 
   wrf@head-cfa.cfa.harvard.edu}}

\begin{abstract}

We present XMM-Newton observations of a trail of enhanced X-ray 
emission extending along the full $8'.7 \times 4'$ region between 
the large spiral galaxy NGC~6872 and the dominant elliptical galaxy 
NGC~6876 in the Pavo Group, the first known X-ray trail associated 
with a spiral galaxy in a poor galaxy group and, with projected length
of $90$\,kpc, one of the longest
X-ray trails observed in any system.  The X-ray surface 
brightness in the trail region is roughly constant beyond 
$\sim 20$\,kpc of NGC~6876 in the direction of the spiral. The 
trail is hotter ($\sim 1$\,keV) than the undisturbed Pavo IGM 
($\sim 0.5$\,keV) and has low metal abundances ($0.2\,\Zs$). The 
$0.5-2$\,keV luminosity of the trail, measured using a
$67 \times 90$\,kpc rectangular region, is   
$6.6 \times 10^{40}$\ergs. 
 We compare the properties of gas in the trail to the spectral 
properties of gas in the spiral NGC~6872 and in the elliptical 
NGC~6876 to constrain its origin. We suggest that the X-ray trail 
is either IGM gas gravitationally focused into a Bondi-Hoyle wake, 
 a thermal mixture of $\sim 64\%$ Pavo IGM gas
with $\sim 36\%$  galaxy gas that has been removed from
the spiral NGC~6872 by turbulent viscous stripping, or both,
due to  the spiral's supersonic motion 
at angle $\xi \sim 40^{\circ}$ with respect to the plane of the sky, 
past the Pavo group center (NGC~6876) through the densest region of
 the Pavo IGM. Assuming $\xi = 40^\circ$ and a filling factor $\eta$
in a cylindrical volume with radius $33$\,kpc and  
projected length $90$\,kpc, the mean electron density and total  
hot gas mass in the trail is 
$9.5 \times 10^{-4}\eta^{-1/2}$\cmc 
and $1.1 \times 10^{10}\eta^{1/2}\Ms$, respectively.
\end{abstract}


\section{Introduction}
\label{sec:introduction}

The study of galaxy interactions with each other and with intragroup
gas in nearby galaxy groups provides a template for understanding
galaxy evolution at high redshift, a time when, according to
hierarchical models, galaxies were rapidly coalescing through accretion
and merger into groups and clusters, enriching, if not producing, the
intragroup medium (IGM), and evolving into their present day 
morphological mix. Physical processes expected to be
important for this evolution form two broad classes: 
(1) tidal interactions such as 
those induced by major mergers (Lavery \& Henry 1988), off-axis galaxy 
collisions (M{\"u}ller \etal 1989), galaxy harassment 
(Moore \etal 1996) or galaxy fly-bys near the core of the 
group/cluster potential (Byrd \& Valtonen 1990), and (2) gas-gas 
interactions, notably ram pressure by the intracluster 
(ICM) or  group IGM gas on the galaxy's interstellar medium 
(ISM), due to the galaxy's motion through the surrounding medium 
(Gunn \& Gott 1972), or ISM-ISM interactions produced in 
galaxy-galaxy collisions (Kenney \etal 1995). These gas-dynamical 
processes may be enhanced by turbulence and viscous effects 
(Nulsen 1982; Quilis \etal 2000) or inhomogeneities and bulk motions 
in the ICM/IGM gas (Kenney \etal 2004). The actions of tidal
forces are identified by the appearance of disturbed 
stellar morphologies, such as stretched stellar tails and folds
(see e.g. Gnedin 2003, Vollmer 2003). 
Observations and simulations suggest 
that minor mergers and off-axis collisions often induce non-axisymmetric
tidal distortions in the central regions of the interacting galaxies, 
initiating gas inflow and starbursts (Kannappan \etal 2004) or other 
nuclear activity. These may cause superwinds and outflow bubbles, 
identifiable by their X-ray signatures (Cecil \etal 2002;
Strickland \etal 2004).

Key observational signatures for ram pressure stripping  
by the ICM/IGM are the appearance of ``cold fronts'' 
(e.g. Vikhlinin \etal 2001; Heinz \etal 2003) and X-ray wakes or tails 
(Stevens \etal 1999; Schulz \& Struck 2001; Acreman \etal 2003).
While these features have been studied in a number of nearby 
galaxies, primarily ellipticals in rich cluster environments, e.g.
NGC~1404 (Jones \etal 1997; Paolillo \etal 2002; Machacek \etal 2004;
Scharf \etal 2004) in Fornax and M86 (Forman \etal 1979; White
\etal 1991; Rangarajan \etal 1995) and NGC~4472 (Irwin \& Sarazin
1996; Biller \etal 2004) in Virgo, wakes and tails
 are low surface brightness structures whose observation is still 
relatively rare. In hot clusters
this is due in part to the rapid depletion of galactic gas 
by the dense ICM near the cluster cores, and also to the 
difficulty of observing a low surface brightness feature  
against the bright cluster background (Acreman \etal 2003). Galaxy groups 
possess a significant, albeit cooler, gaseous IGM
component (Mulchaey \etal 1996, 2003; Ponman \etal 1996;
Osmond \& Ponman 2004). Thus gas-gas 
interactions can occur, with the resulting wakes
and debris trails perhaps easier to study  
(Acreman \etal 2003; Stevens \etal 1999). 
Furthermore, investigation of these systems may illuminate  
the still controversial role of ram pressure in either enhancing or 
quenching star formation within the affected galaxy 
(Fujita 1998; Kenney \etal 2004).

\begin{figure} [t]
\begin{center}
\epsfig{file=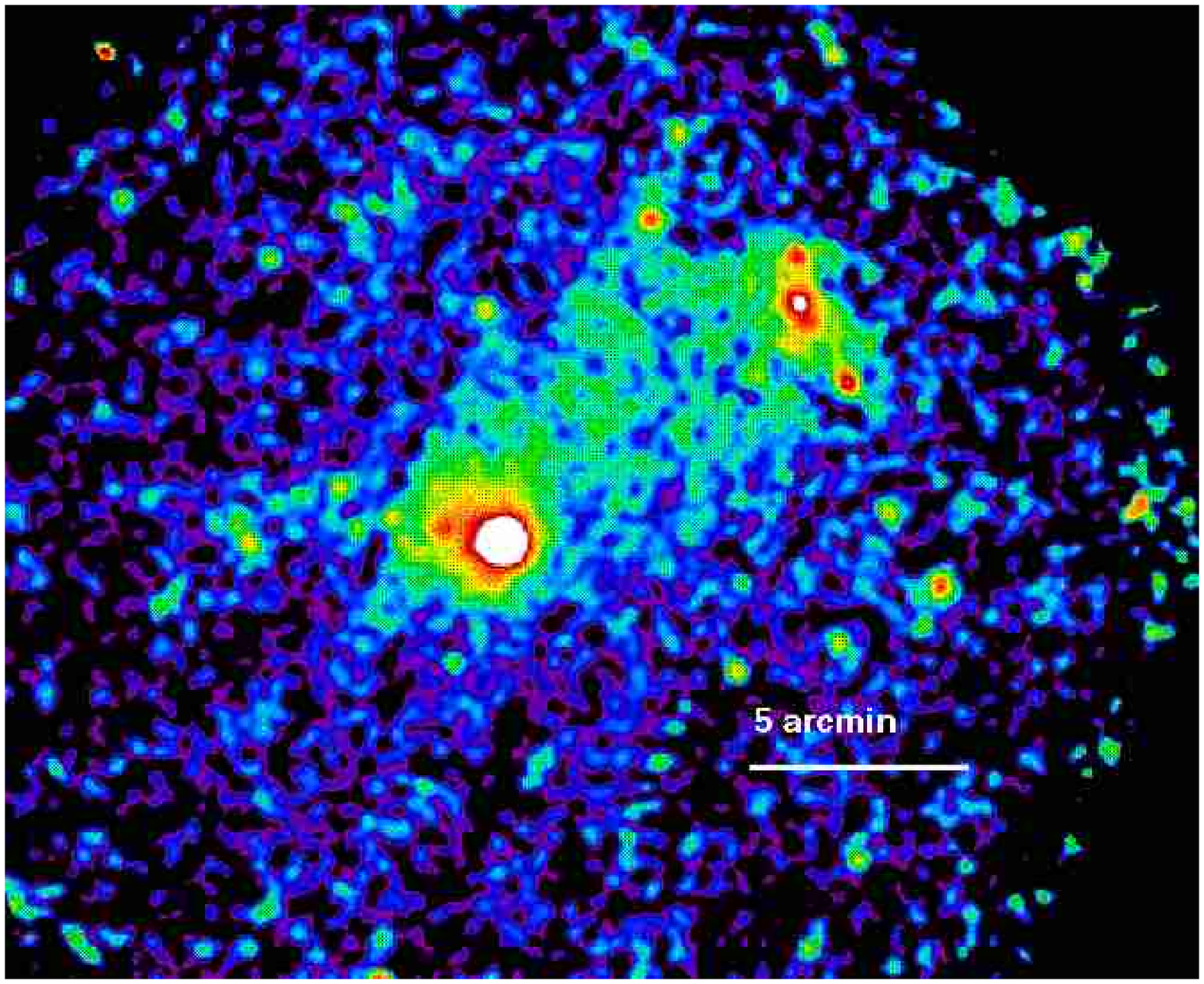,height=3in,width=3in,angle=0}
\hspace{0.2cm}\epsfig{file=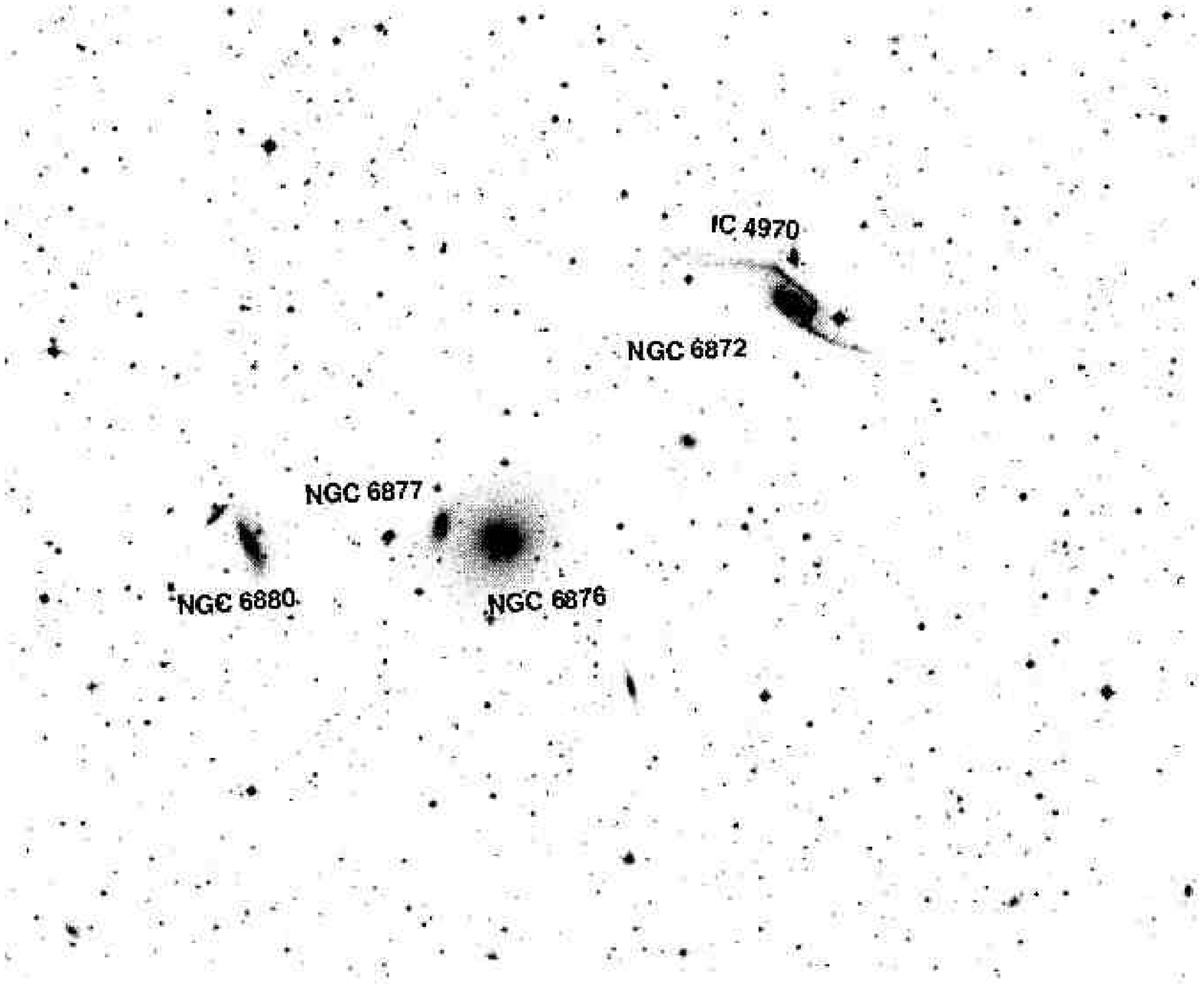,height=3in,width=3in,angle=0}
\caption{(left)The XMM-Newton $0.5-2$\,keV background-subtracted, 
exposure corrected, coadded MOS and PN images of the Pavo Group
showing the X-ray trail of enhanced emission connecting, in
projection, the spiral galaxy NGC~6872 (upper right) to the dominant 
elliptical NGC~6876 (lower left). 
The color scale has logarithmic
stretch from $5 \times 10^{-8}$ to $3 \times 10^{-4}$\nexpunit.
The X-ray trail is evidence of NGC~6872's interaction with the group.
(right) The DSS image of the same 
field matched in WCS coordinates, with the five Pavo galaxies 
(NGC~6872, IC~4970, NGC~6876, NGC~6877, NGC~6880) that are also 
detected in X-rays labeled. Other galaxies in the field are IC 4981 to
the northeast of NGC~6880, IC~4972 to the southwest of NGC~6876, and 
PCG~64439 midway, in projection, between NGC~6876 and NGC~6872.
}
\label{fig:pavo}
\end{center}
\end{figure}
In this paper we present the  XMM-Newton X-ray
observation\footnote{Based on observations obtained with the
XMM-Newton, an ESA science mission with instruments and contributions
directly funded by ESA Member States and NASA} of the 
X-ray trail extending between the  dominant elliptical galaxy 
NGC~6876 and the large spiral galaxy NGC~6872 in the Pavo group. 
This is one of the few known X-ray trails associated with a large spiral
galaxy (see Wang \etal 2004; Sun \& Vikhlinin 2004) and the first 
such  trail observed in a poor galaxy group. As shown in 
Figure \ref{fig:pavo}, the Pavo group, LGG 432 in the Lyon Groups of
Galaxies catalogue (Garcia 1993) with the addition of NGC~6872 and its
companion IC~4970 (Green \etal 1988), is a southern galaxy group comprised
of $\sim 11$ galaxies. The  $18' \times 18'$ Digitized Sky Survey (DSS) 
field (right panel) shows eight of these galaxies   
including the central, dominant elliptical
galaxy NGC~6876 and the giant, spiral galaxy NGC~6872. 
In the left panel of Figure \ref{fig:pavo} we show  $0.5-2$\,keV 
coadded XMM-Newton MOS and PN images.
Before being smoothed by a Gaussian with $\sigma=7''.5$,  
the X-ray images were background subtracted and then corrected for telescope 
vignetting and the spatial dependence of the instrument response 
by use of exposure maps. 
The most striking feature
in this image, other than emission from five individual galaxies, is
the extensive trail of enhanced X-ray emission spanning nearly the 
full $8'.7 \times 4'$ ($\sim 130\,{\rm kpc} \times 60\,{\rm kpc}$) 
projected area between the dominant elliptical NGC~6876 
(near the center of the image) and the large spiral NGC~6872 (upper
right). The image also suggests possible filamentary structure within
the trail, with a reduction in flux by $\sim 28\%$ in the central
region of the trail, near the line joining the two large galaxies. 

Both large galaxies in the field, the dominant elliptical NGC~6876 and the
spiral NGC~6872, have been observed previously in a variety of wavelength
 bands and show evidence for past or ongoing interactions within the 
Pavo group.  The dominant group galaxy NGC~6876 
($\alpha=20^h18^m19.15^s$, $\delta=-70^\circ 51'31''.7$) 
has been observed in the optical, near-infrared (NIR), and mid-infrared
(mid-IR) as well as in X-rays. Optical and NIR Hubble Space Telescope
observations of the inner $5''$ of NGC~6876 show a flat core with 
depressed surface brightness at the center with no evidence for 
dust absorption (Lauer \etal 2002). This has been interpreted as 
evidence that 
a binary black hole resides within $0.''2$ of the center of the
galaxy, a remnant of prior merging activity.
As shown in Figure \ref{fig:pavo}, a smaller elliptical galaxy 
NGC~6877 lies only $1'.4$ in projection to
the east-northeast of NGC~6876. Since 
the relative radial velocity between NGC~6876 and NGC~6877 is  
small ($\Delta v_r \sim 296 \pm 44$\kms, Martimbeau \& Huchra 2004 ),
the two galaxies might form an interacting pair. However, little evidence 
for any  interaction has been found.  In particular, the upper limit on
H$\alpha$ emission in NGC~6876, $L_{{\rm H}\alpha}<2.7 \times
10^{39}$\ergs (Macchetto \etal 1996) indicates little ongoing star
formation at the present time, that might be expected if the galaxies were
interacting. NGC~6876 has been observed
in X-rays by Einstein (Fabbiano \etal 1992; Burstein \etal 1997), 
ROSAT PSPC (O'Sullivan \etal 2001), ASCA (Buote \& Fabian 1998; 
Davis \etal 1999) and XMM-Newton (this work). The left panel of 
Figure \ref{fig:pavo} shows X-ray emission from both NGC~6876 
and NGC~6877, forming, 
in projection, an asymmetric extension to the east. Prior 
X-ray observations lacked the angular resolution to separate these 
two contributions, and it is still unclear whether the common envelope
of X-ray emission is primarily a projection effect or evidence for 
interaction. 

NGC~6872 
($\alpha = 20^h16^m56.48^s$, $\delta=-70^\circ46'5.7''$) 
is a bright, barred spiral galaxy in the 
Pavo group with long, thin, tidally-distorted stellar arms stretching
$ > 4'$ in extent (see Figure \ref{fig:pavo}). It has been observed in
X-rays by Einstein (Fabbiano \etal 1992; Burstein \etal 1997) as well
as by XMM-Newton. H$\alpha$ emission, while absent from the central 
region of the galaxy, is found
concentrated along the spiral arms (Mihos \etal 1993), consistent with 
recent star formation activity. Star formation is concentrated in
regions of the highest measured H$\alpha$ velocity dispersions,
suggesting that it is collisionally induced. From $21$\,cm
observations, Horellou \& Booth (1997) and Horellou \& Koribalski
(2003) showed that $\sim 1.82 \times 10^{10}\Ms$
(for an assumed distance $D=59.8$\,Mpc) of HI gas is distributed 
along and beyond the stellar arms,
avoiding the central bulge and bar regions of the galaxy. 
The extension of HI along the northern arm of the galaxy has a 
``knee'' 
roughly coincident with the location of a 
smaller companion galaxy IC~4970 located $1'.12$ in projection  
to the north of NGC~6872. 
A tidally disturbed stellar bridge between IC~4970  
and NGC~6872, coupled with their small relative radial velocity 
$\Delta v_r = 74 \pm 62$\kms (Martimbeau \& Huchra 2004 ), 
indicates that the two galaxies are interacting. This led to the 
suggestion that an 
off-axis collision may be responsible for both the tidal distortions
and star formation in the arms of NGC~6872 (Mihos \etal 1993; 
Horellou \& Koribalski 2003). Simulations of a prograde, parabolic, 
low inclination collision of IC~4970 with
NGC~6872 (Mihos \etal 1993) successfully reproduced NGC~6872's stellar
bar, the extent of the tidally distorted arms and tidal debris
near IC~4970. However, the simulations required a mass-to-light ratio  
in IC~4970 twice that in NGC~6872. 
The simulations also failed to reproduce the thinness of the spiral
arms, and did not explain why gas was distributed at large radii, thus 
preventing the collision from inducing star formation in the 
bar and nuclear regions.
The elliptical NGC~6876 was  
not thought to contribute to the morphological distortion of 
NGC~6872 (Mihos \etal 1993) because of its $\sim 8$ times larger projected 
distance ($\sim 8'.7$) from the spiral galaxy 
and the significant relative radial velocity 
($849 \pm 28 $\kms, Martimbeau \& Huchra 2004 )
between them.
However, our present analysis using X-ray data from the XMM-Newton satellite 
shows a trail of enhanced X-ray emission,  
extending over nearly the full $8'.7 \times 4'$ region, between
the dominant elliptical NGC~6876 and the large spiral NGC~6872. The
extent of the X-ray trail, as well as the morphology of NGC~6872's
spiral arms, suggests that NGC~6872 interacted strongly 
with NGC~6876 and/or with the group gas during its passage through the 
Pavo group.

In Section \ref{sec:obs} we detail the observation and the 
analysis procedures used. In Section \ref{sec:results} we present our 
main results. In Section \ref{sec:surfacebri} we use  
exposure-corrected X-ray images to compare the surface brightness profiles
of the enhanced X-ray emission (the X-ray trail) between the dominant
elliptical NGC~6876 and the large spiral NGC~6872  with those of  
the undisturbed Pavo IGM. In Section \ref{sec:spec}  
we determine the spectral properties of the Pavo IGM, the X-ray trail,
and the two dominant galaxies, 
NGC~6876 and NGC~6872, and, in Section \ref{sec:dens}, estimate their
mean electron densities and mass contained in hot gas.
In Section \ref{sec:discuss} we interpret
these results arguing that the X-ray trail is most likely either 
gas removed from NGC~6872 by
turbulent viscous stripping (Nulsen 1982) and then thermally mixed with Pavo
IGM gas, IGM gas gravitationally focused into a Bondi-Hoyle wake, or
both. Each process is a result of the supersonic motion of the large 
spiral NGC~6872 past the group center (near NGC~6876) through the 
densest part of the Pavo IGM.
 We summarize our results in Section \ref{sec:conclude}. 
Unless stated otherwise, 
all fluxes are observed fluxes, while all luminosities are
intrinsic, i.e. corrected for absorption.
All uncertainties are $90\%$ confidence limits.
Assuming that the distance to the Pavo group is well represented by the
distance to the dominant elliptical NGC~6876 ($D=53.5$\,Mpc given 
$z=0.01338$, from NED\footnote {http://nedwww.ipac.caltech.edu/}, 
and $H_0=75$\kmsmpc), $1''$ corresponds to a 
physical distance scale of $255$\,pc.
 

\section{Observations and Analysis}
\label{sec:obs}

Our observation consists of a $32.2$\,ks exposure of the Pavo group taken
by the XMM-Newton satellite on 31 March - 1 April 2002.
 The data were obtained using the EPIC camera with MOS1, MOS2 and
PN detectors active, operating in the full-frame mode with thin
filter. The data were analyzed using the standard X-ray processing 
packages: SAS $5.4$, CIAO $3.1$, FTOOLS and XSPEC $11.2$. The data
were cleaned to keep only  standard patterns $\leq 12$ for MOS and 
$\leq 4$ for PN. Bad pixels and columns also were  removed in the 
standard manner. Periods of anomalously high background (flares) were 
identified and removed from the data using light curves for each
detector in the $0.2-12$\,keV and $10-12$\,keV bandpasses, and in 
the $1-5$\,keV bandpass in an outer $11'-12'$ annullus, where the 
effective area is small. 
This reduced the useful exposure times to $17.7$\,ks, $18.8$\,ks, 
and $10.6$\,ks for the MOS1, MOS2, and PN detectors, respectively. 

Backgrounds were subtracted using source free
background event files (Read \& Ponman 2003) provided by the
XMM-Newton Science Operations Center\footnote {see
http://xmm.vilspa.esa.es/external/xmm\_sw\_cal/calib/epic\_files.shtml}
appropriate for the detector, mode and filter with effective exposure
times of $1,055,905$\,s, $1,004,709$\,s, and $351,549$\,s for the
MOS1, MOS2, and PN detectors, respectively. The 
backgrounds were projected onto the coordinates of the current
observation and identical spatial and energy filters were applied to
source and background data throughout, so that the background 
normalization is set by the ratio of exposure
times. This normalizaton was checked by comparing count rates between
the Pavo observation and background files in both the high energy 
($10 - 12$\,keV) band  
and in an outer ($11'-12'$) annular ring at $1-5$\,keV, where
background is expected to dominate. 
We found that the backgrounds differed by $\lesssim 6\%$ for MOS 
($\lesssim 16\%$ for PN) in the $10 - 12$\,keV band and $\lesssim 3\%$
for all detectors in
the $1-5$\,keV band for the outer annular ring. 
We adopt these values as measures of the remaining relative 
uncertainty in the background levels.  Point sources were identified 
in the field shown in Figure \ref{fig:sbmap}, over the $0.5 - 10$\,keV
energy band, using a multiscale wavelet decomposition algorithm set 
with a $5\sigma$ detection
threshold. Nineteen sources were identified, in addition to emission 
from five Pavo galaxies (NGC~6876, NGC~6877,
NGC~6880, NGC~6872, and IC~4970) shown in Figure \ref{fig:pavo}, and
 were excluded from our surface brightness and spectral analyses. 

\section{Results}
\label{sec:results}

\begin{figure} [t]
\begin{center}
\epsscale{0.7}
\plotone{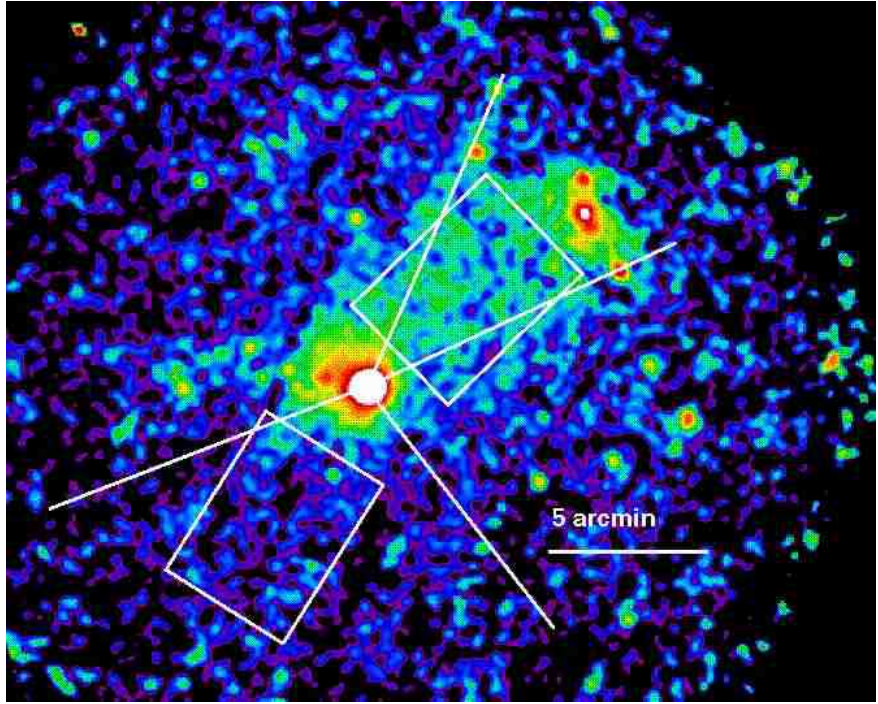}
\caption{The same $0.5-2$\,keV background-subtracted, exposure corrected, 
coadded XMM-Newton images of the Pavo group shown in 
Figure \protect\ref{fig:pavo} overlaid with the 
regions used for analysis. North is up and east is to the left.
Lines outline the northwestern and southern angular sectors used to 
measure the surface brightness profiles, shown in Figure 
\protect\ref{fig:rprof}, of the X-ray trail and Pavo IGM,
respectively. Rectangular regions denote the spectral extraction 
regions for the X-ray trail (northwest) and the Pavo IGM (south). 
}
\label{fig:sbmap}
\end{center}
\end{figure} 
Figure \ref{fig:sbmap} presents the same $0.5 - 2$\,keV coadded 
XMM-Newton images of the central $12'$ circular field of the Pavo Group 
shown in the left panel of Figure \ref{fig:pavo} with regions outlined
for analysis. We construct surface brightness profiles for the MOS and
PN detectors separately, because of their differing effective areas.
We use the angular sectors, shown in Figure \ref{fig:sbmap}, centered on 
NGC~6876 extending from $23.6^\circ$ to $65^\circ$ to the northwest 
for the X-ray trail and from $142^\circ$ to $249^\circ$ to the south for 
the Pavo IGM. All angles are measured clockwise from north. We  
excluded pixels with anomalously low efficiency ($ < 0.45\%$) in order
to mitigate any possible artifical enhancement of low surface brightness
fluctuations due to  poor detector response. 
We also show in Figure \ref{fig:sbmap} the $4'.35 \times 5'.9$ 
($67\,{\rm kpc} \times 90\,{\rm kpc}$) rectangular spectral extraction
regions, for the X-ray trail to the northwest and 
for the Pavo IGM to the south of NGC~6876. 
These two rectangular spectral extraction regions  
are at roughly equivalent off-axis distances in the detectors, 
so that the variation of the telescope effective area over the 
two regions is nearly the same.

\subsection{The Surface Brightness Asymmetry}
\label{sec:surfacebri}

\begin{figure}[t]
\epsscale{0.5}
\begin{center}
\epsfig{file=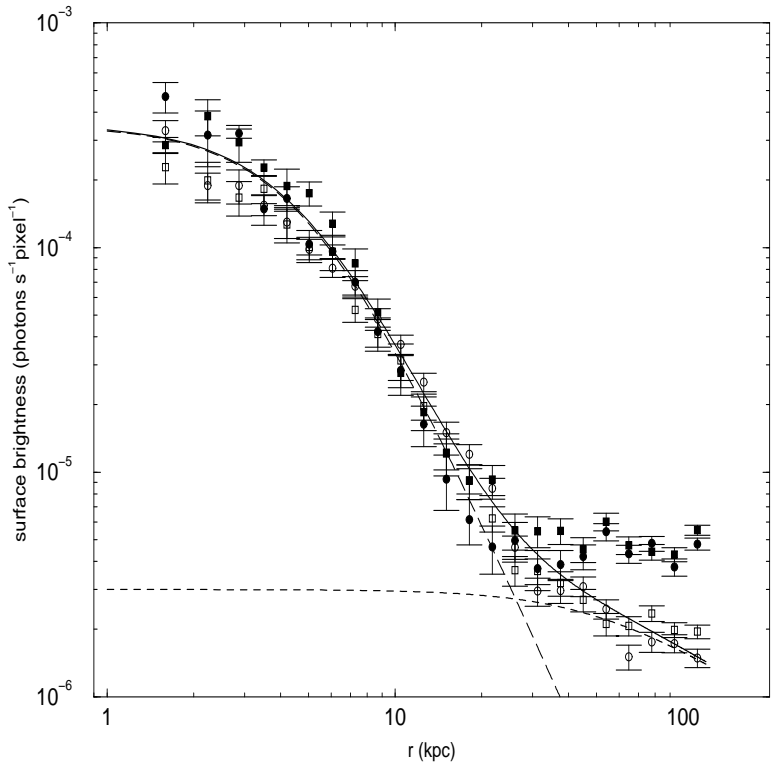,height=4in,width=4in,angle=270}
\caption{Background-subtracted surface brightness profiles 
for $0.5-2$\,keV X-ray emission 
from the elliptical galaxy NGC~6876 through the X-ray trail towards the
spiral NGC~6872 (filled  symbols) and from NGC~6876 through the
undisturbed Pavo IGM to the south (open symbols) using the angular
sectors shown in Figure \protect\ref{fig:sbmap}. Circles denote the 
combined MOS1 and MOS2 data. Squares denote the PN data rescaled by the
ratio of the effective areas of the MOS and PN detectors.
Lines denote the $\beta$-models used 
to describe  the data: a galactic
component for NGC~6876 with core radius $r_c=5$\,kpc and $\beta =
0.65$ (long-dashed line), a Pavo IGM component with 
$r_c =50$\,kpc and $\beta=0.3$ (short dashed line), and the sum of the 
galactic and IGM components (solid line). 
}
\label{fig:rprof}
\end{center}
\end{figure}

In Figure \ref{fig:rprof} we present the background-subtracted, 
$0.5-2$\,keV surface brightness profiles for the X-ray trail 
(filled symbols) and for the Pavo IGM (open symbols) for data from 
the combined MOS detectors (circles) and from the PN detector (squares, 
after rescaling by the ratio of the MOS to PN detector 
effective areas in the $0.5-2$\,keV band),
measured in the angular sectors shown in Figure \ref{fig:sbmap}. 
For the southern sector, the surface brightness profile for NGC~6876, 
is well represented by the sum of
two $\beta$-models, a central galactic (NGC~6876) or group core  
component with  
core radius $r_c=5$\,kpc ($20''$) and $\beta = 0.65$ and an 
extended IGM component with core radius $r_c=50$\,kpc ($196''$) and 
$\beta = 0.3$, out to a radius of $r \sim 120$\,kpc ($\sim 8'$).
Such a two component $\beta$-model characterization of the surface 
brightness profile is expected for galaxy groups 
containing significant amounts of group IGM gas 
(Mulchaey \& Zabludoff 1998). The values we find for the core
radii and $\beta$ indices are in good agreement with those found 
 in recent surveys of   the X-ray emission from similar systems that show
extended X-ray gas and a dominant elliptical at the group center
(Mulchaey \etal 2003; Osmond \& Ponman 2004).
 
The surface brightness profile in the northwest angular sector is clearly
different from the southern sector.  Within $20$\,kpc the profile is
well described by the galactic component ($r_c=5$\,kpc, $\beta =
0.65$) alone, while beyond $20$\,kpc (in the X-ray trail) the surface 
brightness is constant or slowly rising. For projected distances 
$\gtrsim 60$\,kpc from the center of the dominant elliptical NGC~6876, the 
$0.5 - 2$\, keV surface brightness in the northwest region containing the 
X-ray trail is a factor $\gtrsim 2$ larger than that of the
undisturbed Pavo IGM at the same projected distance to the south.

\subsection{Spectral Properties}
\label{sec:spec}

\begin{deluxetable}{ccccc}
\tablewidth{0pc}
\tablecaption{Absorbed Apec Model Fits to the Pavo IGM and the
  X-ray Trail \label{tab:spectrasb}}
\tablehead{
\colhead{Region/Model} & \colhead{MOS1, MOS2, PN} &\colhead{$kT$}  
 &\colhead{$A$} &\colhead{$\chi^2/{\rm dof}$} \\
 & source counts & keV & $\Zs$ & } 
\startdata
 IGM  &$348$, $366$, $926$  & $0.50^{+0.06}_{-0.05}$  & 
   $0.05^{+0.03}_{-0.02}$ & $43.3/39$ \\
 Trail 1T & $788$, $841$, $2000$ & $0.66^{+0.07}_{-0.03}$ &
  $0.06^{+0.01}_{-0.02}$ & $83.6/39$  \\
 Trail 2T & $788$, $841$, $2000$ & $ 0.98^{+0.06}_{-0.07}$ &$0.2 \pm 0.1$  
  &$54.0/39$  \\
\enddata
\tablecomments{ Absorbed APEC model fits to the Pavo IGM and to the debris
field between NGC~6876 and NGC~6872 with fixed Galactic absorption 
 ($4.97 \times 10^{20}$\cms). ``Trail 1T'' denotes a single
temperature APEC model. ``Trail 2T'' 
denotes a two temperature APEC model with one set of parameters fixed
at the IGM model best fit values given in row ``IGM''; while those for
the X-ray trail are free to vary. All fits use the  $0.5-2.0$\,keV bandpass.
}
\end{deluxetable}

To investigate the origin of the enhanced emission in the
X-ray trail between the elliptical galaxy NGC~6876 and the spiral
galaxy NGC~6872, we need to compare the temperature, metal
abundances and density of gas in the X-ray trail to that in 
the undisturbed Pavo IGM and in the galaxies, NGC~6876 and NGC~6872, 
that lie at either end of the X-ray trail.
We extract spectra from circular regions surrounding the two large
galaxies in addition to the spectra from the rectangular regions 
(shown in Figure \ref{fig:sbmap}) for the X-ray trail and Pavo IGM.
For the spiral NGC~6872, we use a $58''.6$ ($14.9$\,kpc) circular
region about the center of 
the galaxy to maximize the count rate, while keeping the region as 
homogeneous as possible. This region includes 
the bulge of the galaxy and inner portions of the northern and
southern spiral arms (see Figure \ref{fig:pavo}), but excludes the 
interacting companion galaxy IC~4970 and extended tidal distortions. 
For the dominant elliptical galaxy NGC~6876, 
we consider two regions, a $62''.7$ ($16$\,kpc) circular region (denoted 
NGC~6876 main) for the central emission from the galaxy, and 
a $99''.7$ ($25.4$\,kpc) circular region (denoted NGC~6876 extended), 
to include the extended envelope of emission to the east-northeast, 
but with a $10''$ ($2.6$\,kpc) circular region that excludes NGC~6877.
For both the northwest region of enhanced emission and the 
southern undisturbed Pavo IGM, the X-ray emission is soft with count 
rates above $2$\,keV consistent with background. We restrict these 
spectral fits to the $0.5-2$\,keV bandpass. 
Spectra for the galaxies may be harder due to either 
unresolved X-ray binaries or nuclear activity. Thus we consider the 
full $0.5-5$\,keV bandpass for
NGC~6876 and NGC~6872, where the MOS and PN  detectors have good 
efficiency and the calibrations are well determined.  For all spectra,
counts are grouped using pre-defined 
groupings resulting in channels of approximately constant 
logarithmic width. The data from the MOS1, MOS2, and PN detectors, for a 
given extraction region, are fit simultaneously using XSPEC 11.2.

\subsubsection{Spectral Properties of the Trail and Pavo IGM}
 
\begin{figure}[t]
\epsscale{0.5}
\epsfig{file=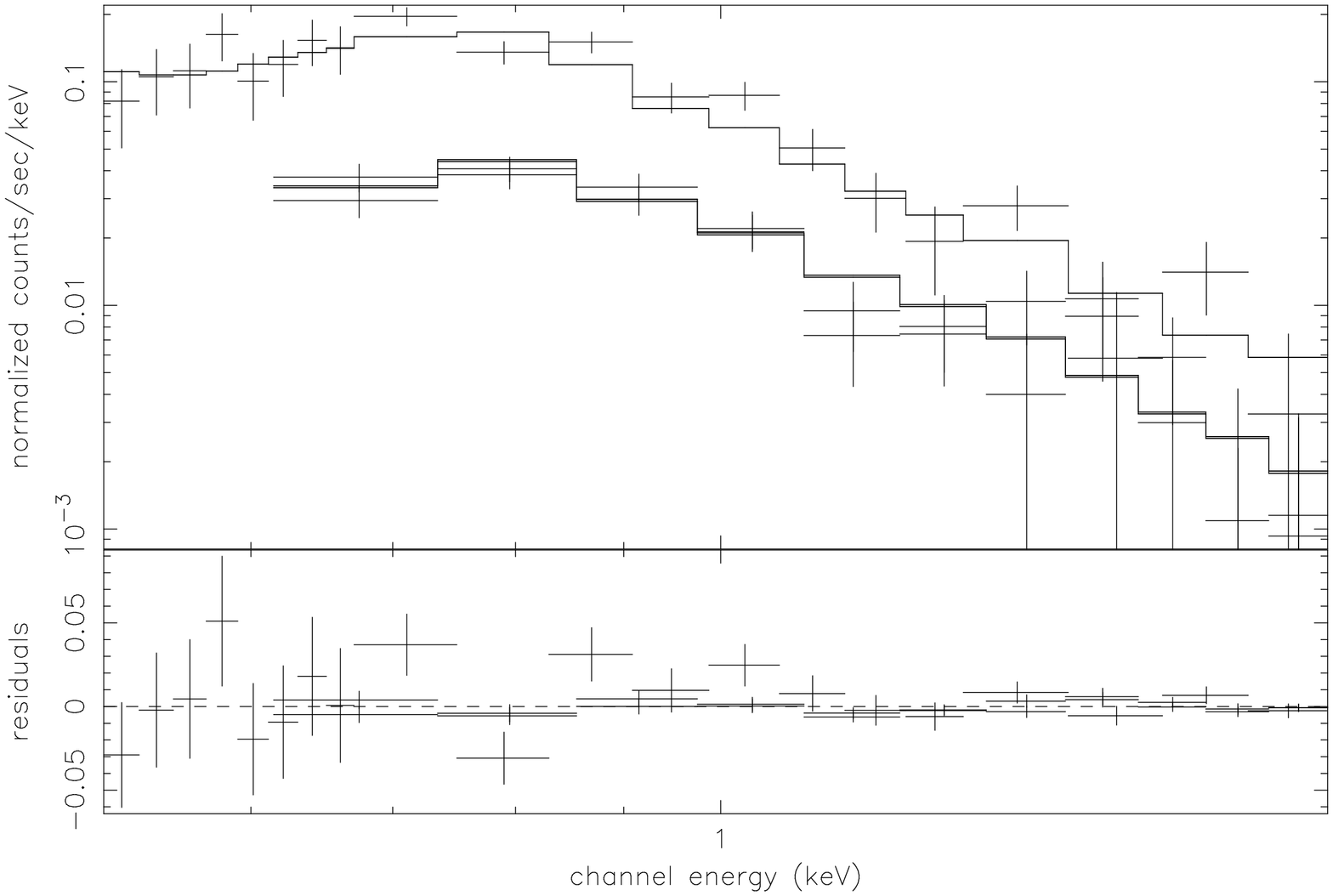,height=3in,width=3in,angle=270}
\hspace{0.3cm}\epsfig{file=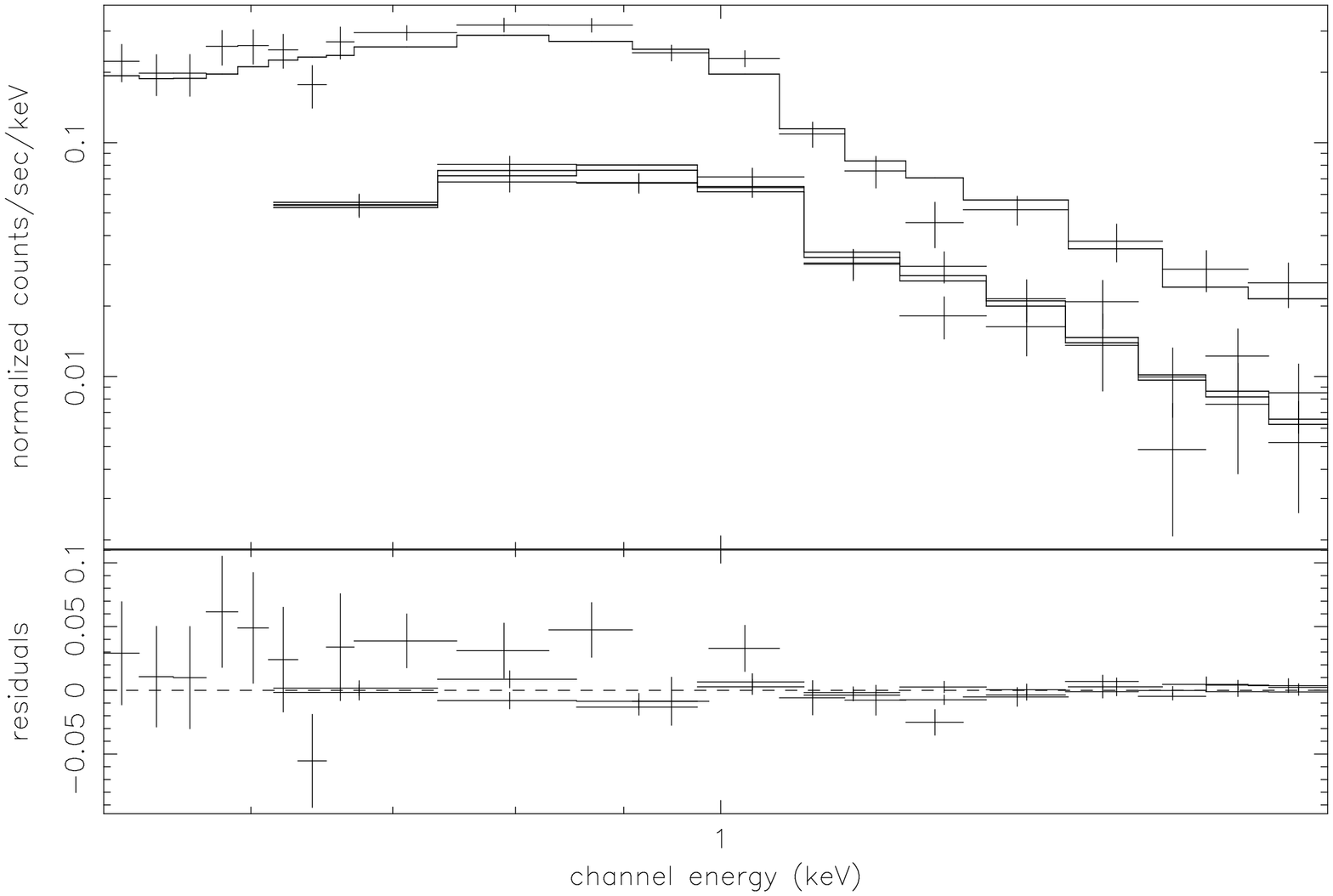,height=3in,width=3in,angle=270}
\caption{(left) Best simultaneous single temperature APEC model fit
to MOS1, MOS2, and PN data for the undisturbed Pavo IGM, extracted from the 
southern rectangular region shown in Figure \protect\ref{fig:sbmap}, with 
temperature $kT = 0.50^{+0.06}_{-0.05}$\,keV, abundance 
$A = 0.05^{+0.03}_{-0.02}\,\Zs$, and fixed Galactic 
absorption $n_{\rm H}=4.97 \times 10^{20}$\cms (NED). 
(right) Best simultaneous two component APEC model fit to the 
X-ray trail extracted from the northwestern rectangular region between
galaxies NGC~6876 and NGC~6872 in Figure \protect\ref{fig:sbmap}. The
X-ray trail component has  temperature 
$0.98^{+0.06}_{-0.07}$\,keV and abundance 
$0.2 \pm 0.1\,\Zs$, while the background IGM 
component is fixed by the IGM model shown in the 
left panel and listed in 
Table \protect\ref{tab:spectrasb}.
}
\label{fig:boxspecsb}
\end{figure}

We consider a variety of possible absorbed APEC plasma models to
describe the spectra for gas in the X-ray trail and Pavo IGM. These  
results are summarized in Table \ref{tab:spectrasb}.  
We are unable to allow all the model parameters to vary freely in 
these fits to the Pavo IGM gas in the southern region, due to 
the faintness of the emission. 
We fix the absorbing column at the Galactic value\footnote {see 
http://heasarc.gsfc.nasa.gov/, Archives \& Software, nH:Column
Density} ($n_{\rm H} = 4.97 \times 10^{20}$\cms), because 
physically  we would not expect excess absorption in the 
undisturbed group gas. Furthermore previous fits to ASCA data for 
the Pavo group (Davis \etal 1999) found an absorbing column consistent
with Galactic.  We find a temperature  $kT = 0.50^{+0.06}_{-0.05}$\,keV
 and metal abundances   $A = 0.05^{+0.03}_{-0.02}\,\Zs$ for the Pavo IGM 
gas in the southern rectangular region. The abundances are in good 
agreement with the results by Davis \etal (1999), who find, using 
ASCA data, $A = 0.09^{+0.36}_{-0.06}$; however, the temperature we find is a
factor $\sim 1.7$ lower. As we will discuss below, their higher
temperature for the Pavo group gas is due primarily to the inclusion
of both the debris trail and the dominant elliptical galaxy NGC~6876
in their measurements. We show our fit to 
the Pavo IGM spectrum in the left panel of Figure \ref{fig:boxspecsb}. 

As shown in the right panel of Figure \ref{fig:boxspecsb}, the spectrum
for gas in the enhanced X-ray emission trail between NGC~6876 and NGC~6872
(northwestern rectangular region in Figure \ref{fig:sbmap}) is   
harder than that of the Pavo IGM to the south.  
A single temperature APEC model
does not provide an acceptable fit to the data. The data are well described 
using a two component APEC model with one set of parameters
(temperature, abundance, and normalization) 
fixed at the best Pavo IGM values from Table \ref{tab:spectrasb} 
to account for background emission
from the Pavo IGM, while the temperature, abundance and
normalization of the second component, modelling the gas in the X-ray  
trail, are allowed to vary. The absorbing column is again fixed at the
Galactic value. The temperature of the gas in the X-ray trail, 
$kT = 0.98^{+0.06}_{-0.07}$\,keV for model ``Trail 2T'', is 
significantly hotter than the IGM gas to the south. Also, the 
abundance $A = 0.2 \pm 0.1\Zs$ is low, marginally consistent with 
that found for the intragroup gas. 

\subsubsection{Spectral Properties of NGC~6876 and NGC~6872}

\begin{figure}[t]
\epsscale{0.5}
\epsfig{file=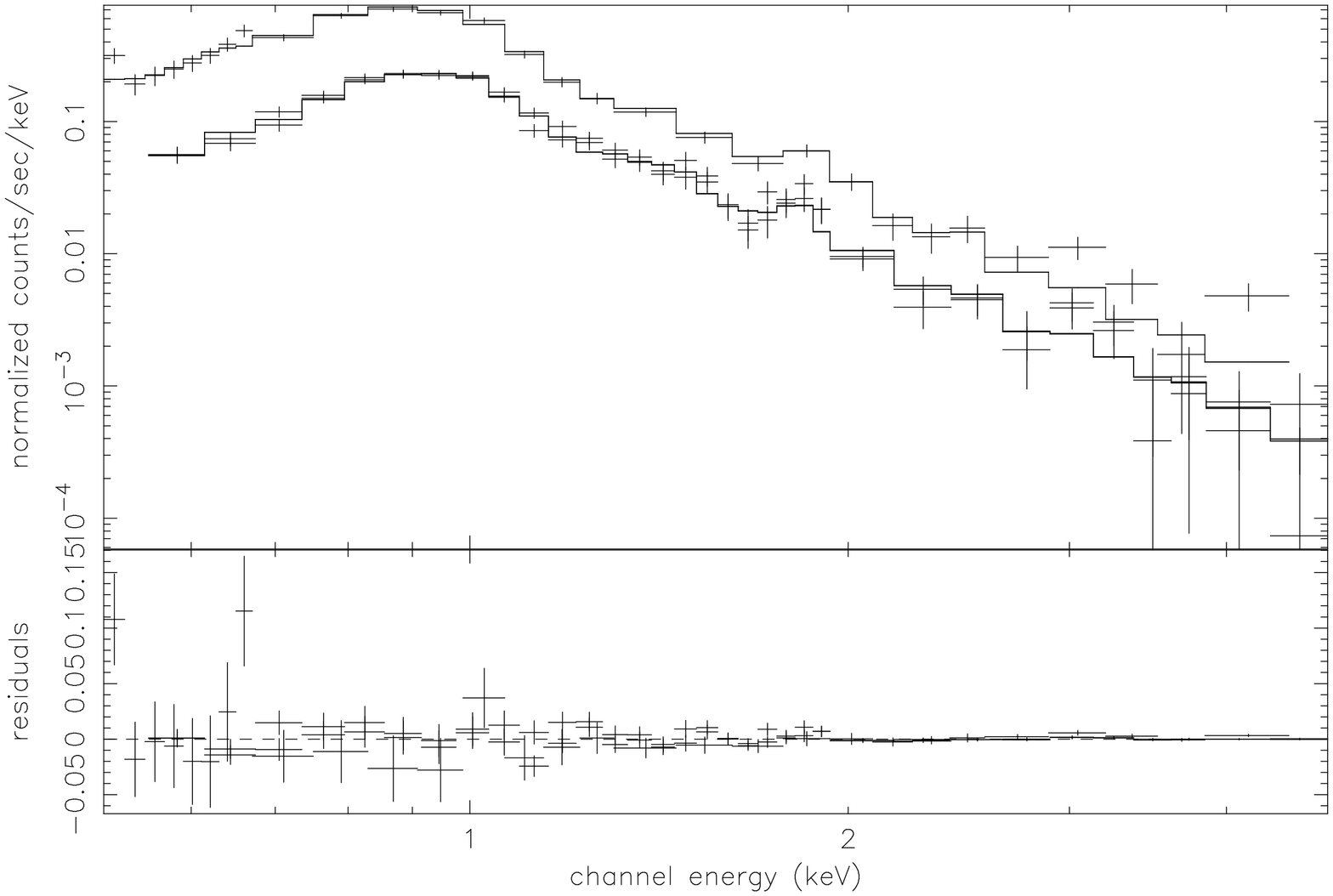,height=3in,width=3in,angle=270}
\hspace{0.3cm}\epsfig{file=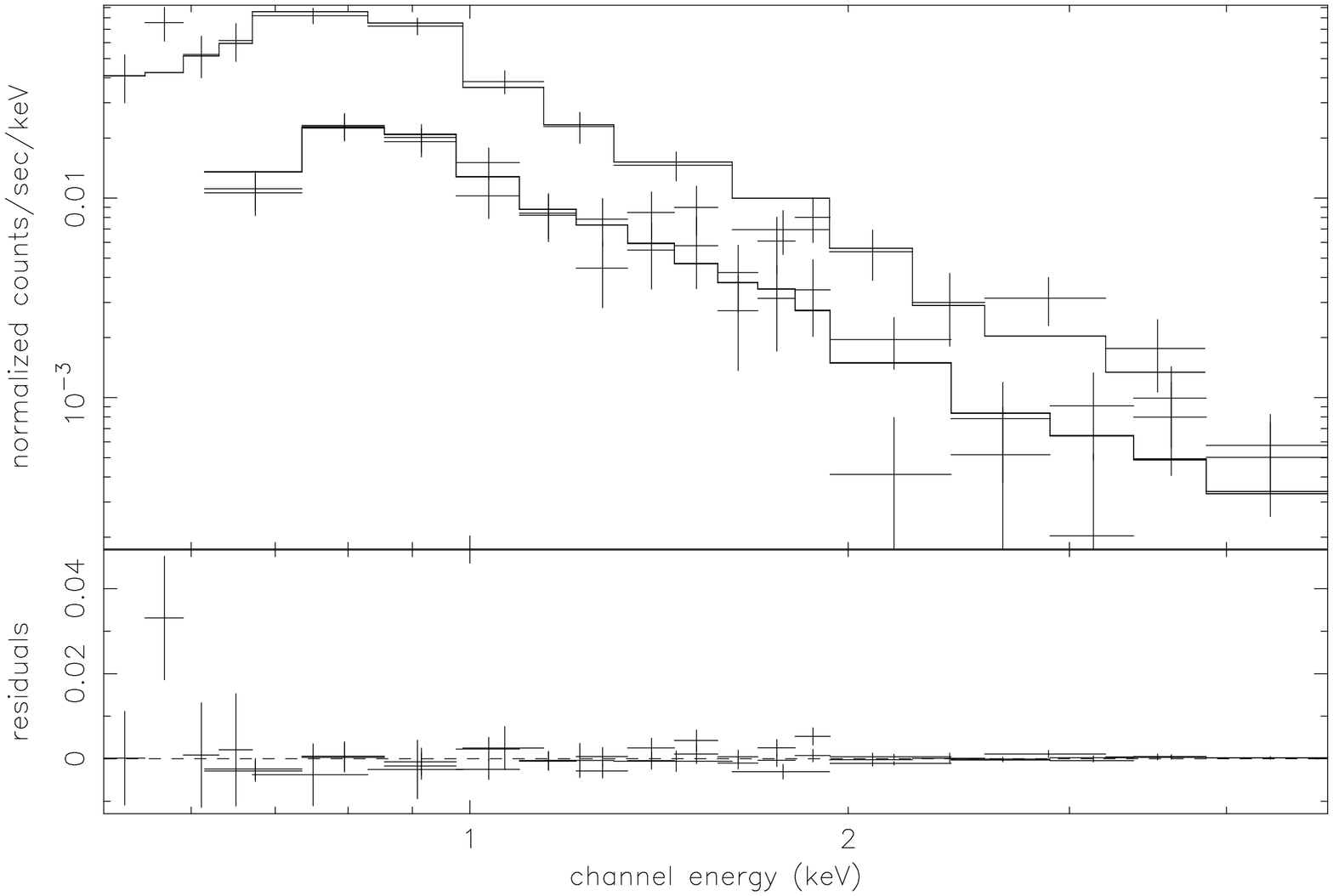,height=3in,width=3in,angle=270}
\caption{ (left) Best simultaneous two temperature APEC model fit
to MOS1, MOS2 and PN data for the extended emission from NGC~6876 in a
circular aperture of radius $99''.7$ with a $10''$ circular region
around NGC~6877 excluded. Absorption is fixed at the Galactic value and the 
abundance is assumed to be the same for both temperature components.
 (right) Best simultaneous two component APEC plus power law model 
fit to MOS1, MOS2 and PN data for a $58''.6$ circular aperture
encircling the central region of the spiral NGC~6872 
with fixed Galactic absorption and solar abundances.
}
\label{fig:galspec}
\end{figure}

\begin{deluxetable}{ccccccc}
\tablewidth{0pc}
\tablecaption{Spectral Fits to NGC~6876 and NGC~6872 \label{tab:galspec}}
\tablehead{
\colhead{Region} & \colhead{MOS1, MOS2, PN} &
\colhead{$\Gamma$} &\colhead{$kT_1$} &\colhead{$kT_2$}  
  & \colhead{$A$}&\colhead{$\chi^2/{\rm dof}$} \\
 & source counts & & keV & keV &$\Zs$ & }
\startdata
NGC~6872 & & & & & & \\
  bulgeA   & $241$, $282$, $594$ & $2.0^{+0.3}_{-0.4}$ & 
  $0.65^{+0.07}_{-0.06}$ & \nodata & $1.0^{\rm f}$ & $48/42$ \\ 
  bulgeB   &$241$, $282$, $594$ & $1.3^{+0.3}_{-0.2}$ & 
  $0.65 \pm 0.05$ & \nodata & $0.2^{\rm f}$ & $45/42$ \\ 
NGC~6876 &  & & & & & \\
 main 1Ta & $1715$, $1923$, $3369$ & \nodata & $0.82 \pm 0.01$ & \nodata 
 & $0.25 \pm 0.03$ & $237/92$ \\
 main 1Tb & $1715$, $1923$, $3369$ & \nodata & $0.77 \pm 0.2$ 
 & \nodata & $0.25 \pm 0.04$ & $179/91$ \\
 main 2T  &$1715$, $1923$, $3369$ & \nodata & $0.76 \pm 0.02$ &
 $2.3 \pm 0.4$ & $1.9^{+1.2}_{-0.5}$ & $107/93$ \\
 extended 2T & $2219$, $2495$, $4377$ & \nodata & $0.75 \pm 0.02$ & 
 $1.6 \pm 0.1$ & $0.95 \pm 0.3$ & $102.3/90$ \\
\enddata 
\tablecomments{ Extraction regions are circular with radii $58''.6$,
$62''.7$, and $99''.7$ for the NGC~6872 ``bulge'', 
  NGC~6876 ``main'', and NGC~6876 ``extended'' models, respectively. 
Superscript ``f'' denotes a fixed parameter. Absorption is fixed at 
the Galactic value ($n_{\rm H}~=~4.97~\times~10^{20}$\cms)
for all but model ``main 1Tb'' where it is fit 
($n_{\rm H} = 1.4^{+0.2}_{-0.3} \times 10^{21}$\cms). 
All spectral fits use the $0.5-5$\,keV bandpass. 
}
\end{deluxetable}
Our spectral fits for the galaxies NGC~6876 and NGC~6872 
are summarized in Table~\ref{tab:galspec}.
A single temperature APEC model with fixed Galactic absorption
(model ``NGC~6876, main 1Ta'') is a poor 
fit to our data for the central $\sim 1'$ of NGC~6876. The fit is 
improved if we allow the absorbing column to vary 
(model ``NCG~6876, main 1Tb'') and is in excellent agreement with the single
temperature model results of Buote \& Fabian (1998) who find
 a temperature $kT \sim 0.88^{+0.15}_{-0.25}$\,keV, abundance 
$A = 0.14^{+0.57}_{-0.07}\,\Zs$, and
absorbing column $n_{\rm H} = 2^{+4}_{-1} \times 10^{21}$\cmc; 
however, the $\chi^2/{\rm dof}$ ($179/91$) for our fit is still large. 
Our data for both the main ($62''.7$) and extended ($99''.8$ with NGC~6877 
removed) circular extraction regions are well represented by a two 
temperature APEC model where the abundances for each
temperature component are constrained to vary together. The 
absorption is again fixed at the Galactic value. 
We find low (high) temperature components
$0.76 \pm 0.02$\,keV~($2.3 \pm 0.4$\,keV) with abundance 
$1.9^{+1.2}_{-0.5}\,\Zs$
for the central $1'$ main region and 
$0.75\pm 0.02$~($1.6 \pm 0.1$)\,keV with abundance 
$0.95 \pm 0.3\,\Zs$ for the extended ($\sim 1'.7$) extraction 
region, respectively. 
Such near solar or super solar abundances are expected in
large elliptical galaxies (Brighenti \& Mathews 1999; Buote 2002). 
Our  results are again in good agreement with
the two temperature Mekal model fits of ASCA data by Buote \& Fabian
(1998) who found temperatures $kT=0.90 \pm 0.14$ and $kT > 1.5$\,keV
with abundance $0.27^{+0.66}_{-0.17}$ for a $2'$ region surrounding NGC~6876
(similar to our ``extended'' region, but including NGC~6877 given 
ASCA's low spatial resolution).
In the left panel of 
Figure \ref{fig:galspec} we show the spectrum and our two temperature 
best fit to the extended ($1'.7$) region of NGC~6876.

In the right panel of Figure \ref{fig:galspec}, we show the spectrum
and best fit model for the central $1'$ (bulge) region of the 
spiral galaxy, NGC~6872, at the northwestern end of the X-ray trail. 
Because of our limited statistics, we cannot constrain all the spectral
parameters in our fits.  We again fix
the absorption at the Galactic value. A simple power law model is 
unable to fit the data, giving $\chi^2/{\rm dof} = 123/44$ for photon
index $\Gamma = 2.7$. A better, but still unacceptable, fit is found
using a thermal APEC model, yielding $kT = 0.75 \pm 0.05$\,keV and 
abundance $0.06 \pm 0.02$ ($\chi^2/{\rm dof} = 81/43$). 
We do find acceptable fits using a power law plus APEC model for both
fixed solar abundance (model ``bulgeA'' with $\chi^2/{\rm dof}=48/42$)
and fixed low abundance $A=0.2\,\Zs$ (model ``bulgeB'' with  
$\chi^2/{\rm dof}=45/42$), suggesting the presence of hot gas. 
 We find the gas temperature ($kT = 0.65$) insensitive to the choice
of abundance over this range; while the photon index for the power law
component in the fits decreases with decreasing abundance from 
$\Gamma \sim 2$ (model ``bulgeA'') to $\Gamma \sim 1.3$ (model
``bulgeB''). This range of values is characteristic of those expected
for unresolved X-ray point sources or a possible AGN (Ptak \etal 1999).

\subsubsection{X-ray Luminosities}

Once the surface brightness and spectral properties of the Pavo IGM,
trail and galaxies have been determined, we can determine their
luminosities. To compare our results for the Pavo group 
with those in recent surveys (Osmond \& Ponman 2004, Mulchaey \etal
2003), we determine the group luminosity by integrating the two 
component $\beta$-model fit to the surface brightness shown in Figure 
\ref{fig:rprof} that includes the central, dominant elliptical galaxy
NGC~6876, but excludes all other sources. We find a $0.5-2$\,keV
luminosity within $r=120$\,kpc (the  extent of the surface brightness
fit) of ${\rm log}\,L_{\rm X}= 41.8$\ergs. 
If we extrapolate our $\beta$-model
fit to the characteristic radius  $r_{500}$,    
\begin{equation}
r_{500} = \bigl ( \frac{124}{H_0} \bigr )\bigl (\frac{T_{\rm
X}}{10\,{\rm keV}}\bigr )^{1/2}\,{\rm Mpc}
\end{equation}
determined from simulations (Evrard \etal 1996)
with $T_{\rm X}$ the temperature in keV and $H_0$ the Hubble 
parameter, we find $r_{500}=0.37$\,Mpc and 
${\rm log}\,L_{\rm X}(r_{500})= 42.5$\ergs. 
This is similar to group luminosities found by Osmond \& Ponman (2004)
who find a mean ${\rm log}\,L_{\rm X}(r_{500})= 42.55$\ergs for their group
sample. However, since our $\beta$-model fit range was restricted to 
$r \lesssim 0.3r_{500}$ and the slopes of $\beta$-models in groups have
been shown to steepen ($\beta \sim 0.4 - 0.5$) at larger 
radii (Osmond \& Ponman 2004), 
our extrapolated luminosity $L_{\rm X}(r_{500})$ may be biased somewhat
high.
 
\begin{deluxetable}{ccccc}
\tablewidth{0pc}
\tablecaption{Observed X-ray Fluxes and Intrinsic Luminosities\label{tab:lumin}}
\tablehead{
\colhead{Region/} & \colhead{Flux ($0.5-2$\,kev)} &
\colhead{Flux ($2-10$\,kev)} &\colhead{$L_{\rm X}$($0.5-2$\,kev)} 
 &\colhead{$L_{\rm X}$($2-10$\,kev)}  \\
component & $10^{-13}$\ergscm & $10^{-13}$\ergscm &$10^{40}$\ergs  & $10^{40}$\ergs }
\startdata
{\bf Trail 2T} &$3.2$ &$0.30$   &$13.3  $  &$1.1$  \\
  \,IGM   &$1.6$   &$0.05 $   &$6.7$ &  $0.2$   \\
  \,Trail  &$1.6$  & $0.25$  &$6.6$ &  $0.9$ \\
{\bf NGC~6872} & & & & \\ 
 bulgeA &$1.2$ &$0.8$ &$4.9$ & $2.9$ \\
  power law   &  & &$2.4$ & $2.8$ \\
  thermal     & & & $2.5$ & $0.07$  \\
 bulgeB &$1.1$ &$1.1$ &$4.6$ & $4.0$ \\
  power law & & &$1.2$ & $3.8$ \\
  thermal   & & &$3.4$ & $0.15$ \\
{\bf NGC~6876} &  & & &   \\
  main 1Tb &$4.5$ &$0.4$ &$24.4$ &$1.6$\\
  main 2T  &$4.5$ &$1.0$ &$17.9$ &$3.8$ \\
  cool comp  & & &$13.6$ &$0.6$ \\
  hot comp & & &$4.3$ &$3.2$ \\
 extended 2T & $5.9$ &$0.96$ &$23.5$ & $3.7$ \\
  cool comp & & & $15.3$ & $0.7$ \\
  hot comp  & & & $8.2$ &$3.0$ \\
\enddata 
\tablecomments{ Fluxes are observed fluxes for the 
region/model listed in Tables \protect\ref{tab:spectrasb} and 
\protect\ref{tab:galspec}.
All luminosities are intrinsic, i.e. corrected for absorption, with 
the absorbing column taken to be Galactic 
($n_{\rm H}~=~4.97~\times~10^{20}$\cms) for all but the model 
``main 1Tb'', where the luminosity is corrected for absorption using 
$n_{\rm H}~=~1.4~\pm~0.2~\times~10^{21}$\cms taken from the fit.
The trail region is rectangular with dimension $ 4'.35~\times~5'.9 $; 
while the regions for the galaxies are circular with 
radii $58''.6$, $62''.7$, and $99''.7$ for the NGC~6872 bulge, 
NGC~6876 main, and NGC~6876 extended models, respectively. Identified
point sources have been subtracted including a $10''$ circular region 
around NGC~6877. 
}
\end{deluxetable}
The observed fluxes and intrinsic luminosities for the
northwestern (trail) region, as shown in Figure \ref{fig:sbmap}, and
the two large galaxies, 
NGC~6876 and NGC~6872, are summarized in Table \ref{tab:lumin} for the
best fit models. The total flux from 
the trail region is  
$3.2 \times 10^{-13}$\ergscm ($0.5-2$\,keV, soft) 
and $3.0 \times 10^{-14}$\ergscm ($2-10$\,keV, hard). Taking 
the distance to the dominant elliptical galaxy NGC~6876
($D=53.5$\,Mpc) as representative of that to the Pavo group,
this implies  an intrinsic soft band (hard band) total luminosity of 
$\sim 1.3 \times 10^{41}$\ergs ($\sim 1.1 \times 10^{40}$\ergs) for 
the region. The expected Pavo IGM background flux and luminosity,
denoted as ``IGM'' in Table \ref{tab:lumin}, are determined from the 
rectangular southern (IGM) region measurements. The IGM emission 
is dominated by soft 
emission, with the ratio of soft to hard band fluxes $\gtrsim~30$.
After subtracting the IGM emission component, the remaining flux    
attributed to the X-ray trail, denoted ``Trail'' in Table
\ref{tab:lumin}, is      
$ 1.6 \times 10^{-13}$\ergscm ($ 2.5 \times 10^{-14}$\ergscm) in the 
soft (hard) bandpass, respectively. The corresponding 
intrinsic soft (hard) band luminosity is 
$6.6 \times 10^{40}$\ergs ($9 \times 10^{39}$\ergs).
While the soft band luminosity from the trail component is 
comparable to that in the IGM component in this region, 
the hard band trail component is five times larger, consistent
with the observed temperatures.

Our measured $0.5-2$\,keV luminosity for the dominant elliptical
galaxy NGC~6876 using the $99''.7$ aperture 
($L_{\rm X} = 2.4 \times 10^{41}$\ergs) is in excellent agreement with 
the ROSAT PSPC results of O'Sullivan \etal (2001) who find
$L_{\rm X} = 2.6 \times 10^{41}$\ergs after
correction to our assumed $53.5$\,Mpc distance. However, the absorbed  
flux we find for the spiral NGC~6872, $1.7 \times 10^{-13}$\ergscm 
(corrected to the Einstein $0.2-4$\,keV bandpass),
 is a factor $\sim 5$ smaller than the $8.66 \times 10^{-13}$\ergscm
found using Einstein data (Fabbiano \etal 1992; Shapley \etal 2001).
This difference is due in part to our focus on the 
central $1'$ of the galaxy,
excluding contributions from IC~4970, resolved point sources and
emission from the outer extensions of the tidally 
distorted spiral arms that would have been included in the 
Einstein measurement. 
As shown in Table \ref{tab:lumin},  
we find an intrinsic $2-10$\,keV luminosity for the
central region of the spiral galaxy NGC~6872  of $2.9 \times
10^{40}$\ergs for $A=1.0\,\Zs$ ($4.0 \times 10^{40}$\ergs for 
$A=0.2\,\Zs$), dominated by the power law component 
of the model. 
Since the X-ray luminosity of low mass X-ray binaries (LMXB's) scales
with the total stellar mass of the galaxy, they contribute 
at most $\sim {\rm a\,few}\times 10^{39}$\ergs to the integrated $2-10$\,keV
luminosity of the spiral NGC~6872 (Gilfanov 2004; Grimm
\etal 2002). On the other hand, if the hard band luminosity is due to 
high mass X-ray binaries (HMXB's), it provides a measure of the 
current star formation rate ($sfr$) in the galaxy
(Gilfanov \etal 2003). Using the $L_{\rm X} - sfr$ relation, 
with our measured $2-10$\,keV luminosity, we predict a star formation 
rate of $4.3\,\Ms\,{\rm yr}^{-1}$ ($6.0\,\Ms\,{\rm yr}^{-1}$),
respectively, for the above two models, consistent with the 
upper limit ($sfr \lesssim 5.6\,\Ms\,{\rm yr}^{-1}$) found by Mihos
\etal (1993) using the $L_{\rm FIR} - sfr$ relation. They argue, 
however, that the low H$\alpha$ luminosity found in NGC~6872 suggests 
that the star formation rate may be much lower. An alternative 
explanation may be that part of the hard band luminosity comes from a 
point source, possibly near the center of the spiral NGC~6872. Such 
ultra-luminous  sources (ULX's) appear to be quite common in galaxies.
Colbert \& Mushotzky (1999) found that $54\%$ of a sample
of $31$ nearby face-on spirals and ellipticals contained such 
near-nuclear point sources, with $0.2-4$\,keV luminosities up to 
$\sim 10^{40}$\,keV. 
We do identify an X-ray  point source in the $0.5-10$\,keV 
band consistent with the optical center of NGC~6872; however because of
the energy dependent broadening of the off-axis point spread function in the
MOS detectors (the radius for $90\%$ encircled energy at $5$\,keV for
a source $8'$ off axis is $\sim 50''$; XMM-Newton User's 
Handbook\footnote{http://xmm.vilspa.esa.es/external/xmm\_user\_support/documentation/uhb/index.html}),
we would expect the hard emission to be distributed over much of the 
extraction region, as is seen.

\subsubsection{X-ray Properties of Other Pavo Galaxies}

\begin{deluxetable}{ccccc}
\tablewidth{0pc}
\tablecaption{X-ray Properties for Other Pavo Galaxies\label{tab:littleguys}}
\tablehead{
\colhead{Galaxy} & \colhead{Net Counts} &
\colhead{Hardness Ratio} & \colhead{$L_{\rm X}$($0.5-2$\,kev)} 
& \colhead{$L_{\rm X}$($2-10$\,kev)} \\
 & $0.5-10$\,keV &    & $10^{40}$\ergs & $10^{40}$\ergs }
\startdata
 NGC 6880 & $68 \pm 20$   & $-0.9 \pm 0.1$  & $0.14$  & $0.02$ \\
 NGC 6877 & $61 \pm 25$   & $-0.7 \pm 0.3$  & $0.13$  & $0.015$ \\
 IC 4970  & $554 \pm 62$ & $0.4 \pm 0.1$  & $2.3$  & $6$ \\
\enddata 
\tablecomments{Net counts are background-subtracted source counts
corrected for telescope vignetting and summed over MOS1, MOS2 and 
PN detectors. Estimated luminosities assume
a $1$\,keV Raymond-Smith plasma model with solar abundance and
Galactic absorption for NGC~6880 and NGC~6877 and an absorbed 
power law model with photon index $1.5$ and absorbing column 
$1.2 \times 10^{21}$\cms for IC~4970, chosen to match the observed
hardness ratios. The distance to the Pavo group,
$D=53.5$\,Mpc, is taken as representative for the galaxies.
}
\end{deluxetable}  
We briefly comment on the X-ray properties, listed in 
Table \ref{tab:littleguys}, of the other three galaxies detected
in X-rays (NGC~6877, NGC~6880, and IC~4970) shown in Figure
\ref{fig:pavo}.
The X-ray emission observed from these three galaxies 
is too faint to formally fit spectra. Instead we measure X-ray
counts in $10''$ ($14''$) circular source regions for NGC~6880 and
NGC~6877 (IC~4970) in the $0.5-2$\,keV and $2-10$\,keV 
energy bands. Local backgrounds are determined from an annular
ring about each source.
We then characterize the emission by computing hardness ratios defined
by $(H-S)/(H+S)$, where $S$ and $H$ are the net source counts observed
in the $0.5-2$\,keV and $2-10$\,keV energy bands, respectively, 
and corrected for telescope vignetting.
We find that the spectra of NGC~6877 and NGC~6880 are soft, with the
hard band 
counts consistent with background (within the large statistical 
errors), yielding  hardness ratios 
$\sim -0.9$, similar to that found in normal galaxies (including
NGC~6876); while emission from IC~4970 is hard
(hardness ratio of $\sim 0.4$), indicative of a nuclear starburst or 
active galactic nucleus (AGN). In Table \ref{tab:littleguys} we 
give luminosities for each source derived using spectra consistent 
with the observed hardness ratios.
The observed hardness ratios for NGC~6877 and NGC~6880 are consistent 
with a $1$\,keV Raymond-Smith plasma model with fixed solar abundance and 
Galactic absorption. Decreasing the temperature to $0.5$\,keV in these 
models decreases the estimated total $0.5-10$\,keV luminosities 
by $\lesssim 10\%$. The hardness ratio for IC~4970 is consistent
with a power law model with photon index $\Gamma \sim 1.5$, typical
of sources with nuclear activity (Ptak \etal 1999), and hydrogen
absorbing column $n_H \sim 1.2 \times 10^{21}$\cms.
Nuclear activity is expected in galaxies
undergoing an off-axis collision, such as the one indicated by the 
tidal stellar bridge between IC~4970 and the northern spiral 
arm of NGC~6872 (seen in Figure \ref{fig:pavo}), as gas from 
the inner disk of the colliding galaxy is driven towards its  center, 
either inducing a starburst or feeding a central black hole 
(Kannappan \etal 2004). If, however, the X-ray emission is due to HMXB's
produced in a recent starburst, the $2-10$\,keV luminosity implies a 
star formation rate of $\sim 9\,\Ms\,{\rm yr}^{-1}$ (Gilfanov \etal
2003). 

\subsection{Electron Densities and Gas Mass Estimates}
\label{sec:dens}

\begin{figure}[t]
\begin{center}
\epsscale{0.5}
\epsfig{file=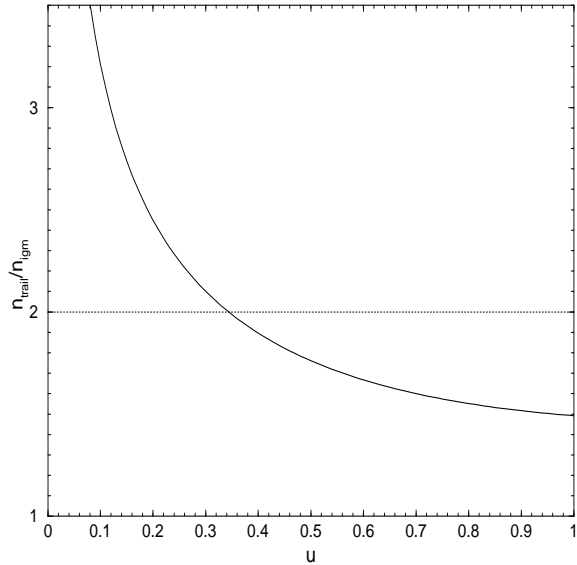,height=3in,width=3in,angle=270}
\caption{Dependence of the ratio $\alpha$ of mean densities for trail
 ($n_{\rm trail}$) and IGM ($n_{\rm IGM}$) gas on the cosine $u$ of the
 angle of motion of NGC~6872 with respect to the plane of the sky. The
 horizontal line denotes $\alpha = 2$.
}
\label{fig:alpha}
\end{center}
\end{figure}

To characterize the X-ray emitting gas in
the trail between NGC~6876 and NGC~6872, we use the spectral
and surface brightness properties to estimate the electron density and
X-ray gas masses in the enhanced emission trail, the undisturbed IGM, 
and the dominant elliptical NGC~6876. We assume that the gas
in the X-ray trail uniformly fills a cylindrical region of radius 
$33.3$\,kpc ($130''.6$) and projected length 
$l_p = 90.2$\,kpc ($353''.6$). The physical length of the trail and,  
thus, the volume occupied by the hot trail gas depend on the cosine of 
the angle ($u={\rm cos}(\xi)$) that the the motion of the spiral
NGC~6872 makes with respect to the plane of the sky. 
We use the XSPEC spectral 
normalization of the $1$\,keV (trail) component to derive a mean 
electron density and estimated total gas 
mass in the trail of $n_{\rm trail} \sim 1.1 \times 10^{-3}
u^{1/2}$\cmc and
$M \sim 10^{10}u^{-1/2}\,\Ms$.  
We determine the IGM density from our fits to the surface brightness 
profile, shown in Figure \ref{fig:sbmap},
using the second (IGM) component of the two component $\beta$-model, that
dominates away from the trail  at distances  
$\gtrsim 35$\,kpc from the center of the elliptical.
We find a central electron density of $1.3 \times 10^{-3}$\cmc for
the IGM component. We then take the IGM electron 
density computed from this $\beta$-model at the center of the trail 
cylinder ($r = 78 u^{-1}$\,kpc from NGC~6876) as representative of 
the mean electron density of the undisturbed IGM in that volume. 
In Figure \ref{fig:alpha} we plot the ratio $\alpha$ of the mean gas density 
in the trail ($n_{\rm trail}$) to that in the IGM ($n_{\rm IGM}$) 
 as a function of $u$. For all but the largest angles ($\xi >
70^\circ$), $\alpha \lesssim 2$ such that the overdensity in the trail
is $\lesssim 1$.

For the elliptical galaxy NGC~6876, we use the galaxy component of 
the $\beta$-model fit to the surface 
brightness profile in Figure \ref{fig:sbmap}, that dominates at small
radii, to calculate the central electron density and integrated gas
mass within $16$\,kpc (NGC~6876 main) and $25.4$\,kpc (NGC~6876
extended) spherical volumes. 
We find a central electron density of $\sim 2 \times 10^{-2}$\cmc 
and hot gas masses of $\sim 1.7 \times 10^9\Ms$ ($\sim 3.4 \times 10^9\Ms$) 
within the central $16$\,kpc (extended $25.4$\,kpc)
regions. However, the  errors on these estimates are
large, due to uncertainties in the central surface
brightness caused by our limited statistics and to uncertainties in
the gas metal abundance that influence the value we adopt for the 
$0.5-2$\,keV emissivity  
($\Lambda \sim 1.9 \times 10^{-23}$\,erg\,cm$^{3}$\,s$^{-1}$ ). 

\section{On the Nature and Origin of the Trail}
\label{sec:discuss}

The observation of a wake or trail behind a galaxy is one of the few 
ways known to determine the direction of a galaxy's motion in the plane 
of the sky and thus constrain its dynamical motion through the group
or cluster. In this case the direction and extent of the trail 
between NGC~6876 and NGC~6872, shown in Figures \ref{fig:pavo} and 
\ref{fig:sbmap}, indicate that  the spiral
NGC~6872 has recently accomplished a fly-by of the dominant 
elliptical NGC~6876. Since the speed of sound in the $0.5$\,keV 
Pavo  group gas is $\sim 365$\kms 
and assuming the central elliptical 
NGC~6876 is at rest relative to the IGM, 
the relative radial velocity 
($v_r = 849 \pm 28$\kms, Martimbeau \& Huchra 2004)
between NGC~6872 and NGC~6876 suggests that the motion through the 
group gas is supersonic (Mach $M \gtrsim 2.3$). Key observations that 
need to be explained by any model are the length of the trail and the 
density, temperature and abundance of gas in the trail relative to
that of the surrounding IGM. 

Four possible physical mechanisms 
might contribute to the formation of the observed X-ray trail during
such a fly-by event. The X-ray trail might be : 
\begin{itemize}
\item gas tidally stripped from either NGC~6876 or 
NGC~6872, during the close encounter of the two systems. 
\item IGM material gravitationally focused into
a Bondi-Hoyle wake (Bondi \& Hoyle 1944; Bondi 1952; Hunt 1971;
Ruderman \& Speigel 1971) behind NGC~6872, as the IGM gas flows 
past the large spiral. 
\item galaxy ISM stripped from the spiral NGC~6872
by ram pressure (Gunn \& Gott 1972), as the galaxy passed 
through the densest part of the Pavo IGM. 
\item gas stripped from NGC~6872 by 
turbulent viscosity (Nulsen 1982), due to the supersonic motion of the
spiral through the Pavo IGM, that is then thermally mixed with the
ambient group gas.
\end{itemize}
Identification of the dominant process in the formation of 
the trail may not be easy, since all of these processes may act
together to some degree during the passage of the spiral galaxy 
through the core of the group. We discuss each of these mechanisms below.

\subsection{Tidal Interactions}
\label{sec:tides}

Tidal interactions between the spiral NGC~6872 and the dominant 
group elliptical NGC~6876 during the fly-by, not just between the 
spiral and its small companion galaxy IC~4970 (Mihos \etal 1993; 
Horellou \& Koribalski 2003), may well have contributed 
to the extended stellar tidal tails 
in NGC~6872. These stellar tidal tails stretch  $\sim 2'$ 
($\sim 30$\,kpc) to the east and west of its bulge 
(see Figure \ref{fig:pavo}) and
have associated with them $\sim 1.4 \times 10^{10}\Ms$ 
(distance corrected to $53.5$\,Mpc) of HI gas  
 (Horellou \& Booth 1997; Horellou \& Koribalski 2003). 
However, since tidal interactions affect gas and stars similarly and
tend to be less effective for high speed encounters, they
are unlikely to be the dominant process at work in the formation of the 
X-ray trail. If the $\sim 10^{10}\Ms$ of hot gas in the X-ray trail 
originated in NGC~6872, it would represent more than $40\%$ of the original
gas mass in the galaxy. It is unlikely that tidal forces would
displace that much hot gas without there also being evidence in the 
stellar or HI gas distributions of stars or cool gas displaced toward 
the region of the X-ray trail. Similarly, the gas cannot have come 
from the elliptical NGC~6876 in the interaction. The mass of gas in 
the X-ray trail is a factor $\gtrsim 3$ greater than the hot gas mass 
retained (within $25$\,kpc) by the dominant elliptical NGC~6876. 
If tidal interactions were that efficient in removing gas from the
galaxy, one would expect similar distortions in the distribution of
the stars, that are not seen. Also the heavy element abundance of 
gas in the trail is much lower than that measured in NGC~6876.

\subsection{Bondi-Hoyle Wake}
\label{sec:wake}

A second possibility is that the trail is a Bondi-Hoyle wake, where 
the group IGM is adiabatically compressed and heated, due to the   
gravitational focusing of the surrounding IGM by NGC~6872's 
gravitational potential, during the passage of 
the spiral through the group core.  
Qualitatively the fact that the 
measured heavy element abundance of gas in the 
trail is low, consistent with that of the Pavo IGM, rather than the 
near solar or super solar abundances expected for galactic gas or gas
from a superwind, favors the interpretation of the trail as a 
wake, consisting primarily of IGM material. The observed density in 
the trail (see Figure \ref{fig:alpha}) is 
also reasonable, since adiabatic compression can readily produce 
overdensities of order unity.  

Of more concern are the dimensions of the tail, i.e. 
whether gravitational focusing by the spiral can produce significant
density perturbations that extend for $\gtrsim 90$\,kpc behind the 
galaxy.  The characteristic length scale
for gas to be drawn into the wake is given by the Bondi-Hoyle
accretion radius $R_A$, whose analytical form depends
critically on whether the relative velocity of the galaxy 
$v$ with respect to the surrounding medium is subsonic, where  
adiabatic gas infall dominates, or highly
supersonic, where the accretion process is dominated by the 
dynamical motion of the gas (Bondi 1952). In the extreme subsonic 
regime ($v \ll s_{\rm IGM}$, where $s_{\rm IGM}$ is the speed of sound
in the surrounding medium), the accretion radius is given by 
$R_A = 2GM_g/s_{\rm IGM}^2$, where $G$ is the gravitational 
constant and  $M_g$ the accreting galaxy's (total) mass, while for 
hypersonic velocities ($v \gg s_{\rm IGM}$) $R_A = 2GM_g/v^2$.  
Previous analyses at intermediate velocities appropriate for
galaxies moving in groups or clusters have either used numerical 
simulations of highly idealized accretors, i.e.
a gravitating point source (e.g. Bondi 1952, Hunt 1971) or totally 
absorbing sphere (e.g. Ruffert 1994), or have limited the accretors to
slowly moving systems (Mach number $M \leq 1$), where the size of the 
galaxy is smaller than the accretion radius (Sakellou 2000). These 
results are not directly applicable to the trail we observe. 

\begin{figure}[t]
\begin{center}
\epsscale{0.5}
\epsfig{file=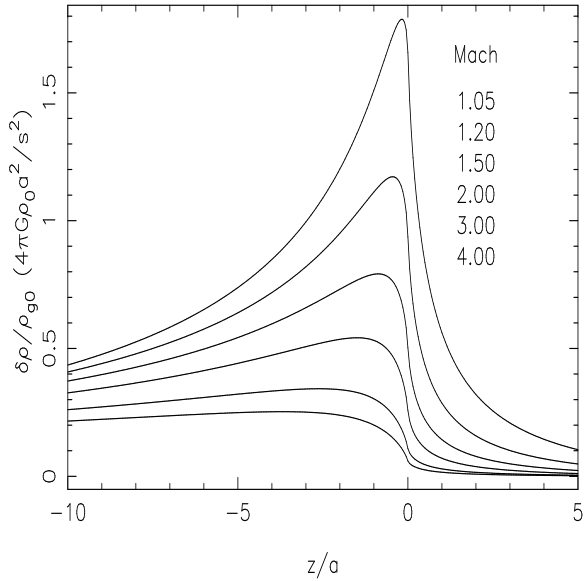,height=3in,width=3in,angle=270}
\caption{Density perturbation on the accretion axis 
produced by an NFW potential with central density $\rho_0$ and 
inner radius $a$, moving supersonically through a uniform medium 
with sound speed $s$. Negative values of $z/a$ indicate distances 
along the accretion axis downsteam of the perturber. 
}
\label{fig:bondi}
\end{center}
\end{figure}

We can obtain an exact solution to the   
linearized flow equations for adiabatic motion, given in the Appendix,
 by convolving the 
Greens function for the equations, i.e. the solution for a gravitating 
point source moving supersonically through a uniform medium found by 
Ruderman \& Spiegel (1971),  with the density distribution  
for more realistic galaxy potentials. In the 
currently favored cosmologies, simulations show that the 
gravitational potential for galaxies is dominated by dark matter. 
Thus the scale for Bondi accretion is set by the mass 
distribution of the dark matter halo, whose density distribution, even
for spirals, is well represented by the spherically symmetric NFW form
\begin{equation}
\rho (r) = \frac {\rho_0}{\frac {r}{a}\bigl ( 1 + 
    \frac{r}{a} \bigr )^2} 
\label{eq:nfw}
\end{equation}
where $\rho_0$ is the central dark matter density and $a$ is the inner 
NFW radius (Navarro, Frenk \& White 1995, 1996, 1997). In the Appendix
we solve the linearized flow equations for  
an NFW gravitational potential, i.e. a gas free galaxy, 
moving supersonically through a uniform medium with sound speed $s$. 
In Figure \ref{fig:bondi}, we show these solutions for the density
perturbation $\delta\rho/\rho_{0g}$ on the accretion axis for Mach 
numbers $1.05 \leq M \leq 4$. Negative (positive) values for 
$z/a$ denote downstream (upstream) values of the distance from the 
perturber along the accretion axis.
Several features of Figure \ref{fig:bondi} are noteworthy. First 
we see a  strong asymmetry between the upstream and downstream values 
for the density perturbation, producing a wake of adiabatically 
compressed gas downstream of the galaxy.
Secondly,  as the speed of the halo through the medium increases,
the peak of the density distribution moves somewhat downstream and 
its magnitude dramatically decreases, such that for the large 
Mach number ($M > 2.3$) expected for NGC~6872, the pile-up of 
accreted material near the center of the galaxy, predicted for 
subsonic motion (Sakellou 2000), does not occur. This is 
consistent with the low X-ray luminosities observed in the central 
region of the spiral. Third, as the Mach number increases
the magnitude of the perturbation decreases with downstream distance 
much more slowly than for lower galaxy speeds relative to the medium.
For example, for Mach $M \sim 3$ the density perturbation peaks at a 
downstream distance $z \sim 1.5a$ and decreases by $\lesssim 20\%$  
at $z \sim 10a$. Thus for sufficiently large halos and high speeds, 
the density perturbation can be substantial at large downstream 
distances, producing a long wake comparable to what is seen. 

Massive dark matter halos with large virial radii may be common for spirals. 
For example, dynamical models for the motion of gas in the Magellanic 
Streams associated with our Galaxy indicate a dark matter halo with  
radius $> 100$\,kpc (Binney \& Tremaine 1987). 
Such a large halo may be plausible for
NGC~6872, especially if it is the center of a spiral-dominated
subgroup moving through Pavo. To estimate 
the size of the density perturbation expected for the motion of  
NGC~6872 through the Pavo 
IGM, we rewrite the scale factor for the perturbations in 
Figure \ref{fig:bondi} in terms of observable quantities, i.e. 
\begin{equation}
\frac{4\pi G\rho_0 a^2}{s^2} = 
   \frac{v^2_{\rm circ}}{0.216 s^2_{\rm IGM}}
\label{eq:bondiscale}
\end{equation}
where $v_{\rm circ}$ is the (maximum) rotation velocity for the spiral
and $s = s_{\rm IGM}\sim 365$\kms 
is the sound speed in the undisturbed 
Pavo IGM. Although the rotation curve for NGC~6872 has not been
measured, we can estimate the rotation velocity from the 
I-band or H-band Tully-Fisher relations (Pierce \& Tully 1992). Using 
 I-band (H-band) magnitudes
$11.2$ ($8.64$) from NED and distance $D = 53.5$\,Mpc to NGC~6872, we 
find a rotation velocity $v_{\rm circ} \sim 230$($350$)\kms 
 for the spiral. We caution the reader that the Tully-Fisher
relation has not been calibrated in interacting systems like NGC~6872,
so that optical measurements  of the rotation curve in that system
are needed to fully understand the kinematics and 
constrain the gravitational potential. However, the circular 
velocities we infer from the Tully-Fisher relation, 
$230 \lesssim v_{\rm circ} \lesssim 350$\,kms are reasonable, 
comparable to those observed in more normal large spirals. 
Using the I-band result as a conservative estimate for 
the circular velocity, we determine the other NFW parameters
for the spiral NGC~6872 by comparison with simulations (Bullock \etal
2001). We find a virial mass for NGC~6872 of $\sim 3 \times 10^{12}\,\Ms$, 
the present epoch NFW concentration
parameter  $c \sim 15$, and the NFW inner radius $a \sim 20$\,kpc. 
Using these parameters for Mach $M = 3$ motion in Figure
 \ref{fig:bondi}, the maximum IGM density perturbation (on the
accretion axis) in the trail is estimated to be $\sim 0.6$ dropping
to $\sim 0.5$ at $z \sim 10a \sim 200$\,kpc, comparable to the observed 
overdensity and length of the trail for angles 
$\xi \lesssim 45^\circ$ (see Figure \ref{fig:alpha}).   

A major concern is the effect such a large halo 
might have on the elliptical NGC~6876 
during the close encounter of the two systems. Given the inferred 
large virial mass of the spiral NGC~6872, the halos of the two 
galaxies would most likely still be overlapping, yet the stellar and 
gas distributions of the elliptical do not appear strongly distorted. 
Similarly, it is unclear, without better numerical simulations, whether 
such a large halo for NGC6872  would prevent the formation of its  
observed tidally-extended spiral arms either through interaction with 
the core group potential or with its companion, IC~4970. However, 
as indicated in Figure \ref{fig:bondi}, the perturbation remains 
substantial downstream even for smaller $a$, that might result if the 
dark matter distribution for the galaxy has been modified by 
interaction with the group potential. 

Another concern is the temperature rise observed in the trail. 
For the above overdensities of $\sim 0.6$, conservation of entropy
predicts a temperature ratio $T_{\rm trail}/T_{\rm IGM} \sim 1.4$, 
much less than that observed. To produce the observed ratio, 
 $T_{\rm trail}/T_{\rm IGM} \sim 2$, would require  
a density ratio $n_{\rm trail}/n_{\rm IGM} \sim 2.7$ and 
overdensity $1.7$. Using 
our observed values for the density ratio in Figure \ref{fig:alpha} 
to fix the kinematical parameters for NGC~6872, we find an angle 
$\xi \sim 80^\circ$ and thus Mach number $M \sim 2.4$. From Figure 
\ref{fig:bondi} we see that an 
overdensity $\sim 1.7$ is possible provided the circular velocity 
of NGC~6872 is large ($\sim 350$\kms) and/or nonlinear effects 
become important, both assumptions testable in simulations, but  
 then the inferred length of the observed trail ($\sim 500$\,kpc for  
 projected length $\sim 90$\,kpc) is probably 
too long. Thus it is unlikely that Bondi accretion alone 
can account for the properties of the observed trail. However, it 
may act in concert with other physical processes that can heat the 
gas in the trail.

Key observational signatures, detectable in deeper 
Chandra and XMM-Newton X-ray exposures, which are capable of
distinguishing between gravitational focusing of the IGM into a 
Bondi-Hoyle wake and  competing explanations for the trail, are the 
trail gas metal abundance and the temperature and density profiles 
along the trail. Since, in the Bondi-Hoyle wake scenario, the trail 
is composed only of gravitationally focused IGM gas, the trail gas 
abundance should match that in the undisturbed IGM. Entropy 
conservation also implies that the temperature and density profiles 
are correlated. The temperature of 
the trail is always hotter than the ambient IGM, due to adiabatic 
compression, with the peak temperature occurring near the spiral 
NGC~6872 where the density perturbation is greatest.

\subsection{Ram-Pressure Stripping}
\label{sec:ram}

A third possible explanation for the X-ray trail in this system 
is that it is a combination of Pavo IGM gas and ram-pressure-stripped 
ISM gas from NGC~6872 that has been shock heated and
focused into the trail of enhanced X-ray emission (or wake)
behind the spiral NGC~6872, due to the supersonic motion 
of the galaxy through the group IGM, during the spiral's 
initial infall through the Pavo group center (near the dominant 
elliptical NGC~6876 and through the densest part of the Pavo IGM).
To better understand whether this explanation is plausible, we compare
our measured properties for gas in the X-ray trail to the results of 
simulations by Stevens \etal (1999) and those by Acreman \etal
(2003) for galaxies moving supersonically through ICM gas. 
The simulations use a two dimensional PPM
implementation of hydrodynamics with gravity and a simple model for 
mass replenishment from supernovae, to study the dependence of
observables on cluster gas  temperature, Mach number, and galaxy mass 
replenishment rate for a canonical elliptical galaxy of total mass 
$1.2 \times 10^{12}\Ms$ moving at constant
velocity through the cluster gas (Stevens \etal 1999),  and on the 
galaxy gas halo size, galaxy
mass, galaxy mass replenishment rate and infall radius for radial
infall of a spherical galaxy through $2.7$\,keV cluster gas 
(Acreman \etal 2003). Our data are most similar to the cool ($1$\,keV)
cluster simulations of Stevens \etal (1999). 
While the effects of the very different geometrical cross-section 
presented by the gas distributions in the spiral NGC~6872  
 are difficult to predict and require further simulation,  
one might expect that the gross 
properties of the wakes would be similar.  

The three potentially visible features of a galaxy undergoing ram-pressure 
stripping, due to supersonic motion through the surrounding cluster 
gas, are a bow-shock preceeding the galaxy, a cold front at the galaxy's
leading edge, and an extended tail or wake trailing the galaxy. 
Of these, the bow-shock is the least visible and, in our case where the
spiral NGC~6872 is observed far off axis in the detector, would not 
be expected to be seen. The trailing wake is the next most visible 
feature found in the simulations (Stevens \etal 1999; Acreman \etal 2003).
Stevens \etal (1999) find that the most visible, densest wakes form
behind galaxies in cool clusters, with high mass replenishment rates; 
while Acreman \etal (2003) find the wakes are most visible for galaxies 
with substantial initial gas content on their first passage through 
the cluster.  Both conditions may hold in our case, since 
the Pavo group is cool, the trail morphology suggests a fly-by through
the group center, and NGC~6872 has substantial HI gas along its
tidally extended spiral arms.

Simulations by Stevens \etal (1999) suggest that 
ram-pressure stripping of the ISM in galaxies in cool clusters or 
groups may not be efficient, with mass retention factors between 
$30\%$ for low mass replenishment rates, appropriate for galaxies with 
only an old population of stars, to  $75\% -89\%$ for higher mass 
replenishment rates more appropriate for galaxies with recent star 
formation. Recent star
formation is observed in and at the ends of the tidally distorted 
spiral arms of NGC~6872 (Mihos \etal 1993), such that the latter 
case with less efficient stripping may be more applicable. 
Given that the mass of hot gas in the 
X-ray trail is $\sim 40\%$ of the HI 
mass measured in NGC~6872, this implies a mass retention factor of 
$\sim 60\%$, in agreement with the simulations. 
Stevens \etal (1999) also argue that the temperature of the 
wake can be hotter or cooler than the ambient cluster gas, depending 
on whether the wake consists primarily of shocked IGM gas or cooler 
ram-pressure stripped ISM gas. Our hotter temperature for the trail
would imply that shock-heated IGM gas dominates, consistent with the 
observed low abundance ($\sim 0.2\,\Zs$) of the trail gas. However, 
the small range of temperatures reported for their simulated wakes  
were all comparable to (within $\sim 10\%$) or less than the
temperature of the surrounding ICM,
rather than the factor $\sim 2$ higher temperature we observe.

The simulations of Acreman \etal (2003) provide insight into the 
expected luminosity and extent for trails produced by ram-pressure 
stripping. They find that the broadest and
brightest wakes occur for galaxies undergoing their first core
crossing of the cluster with simulated $0.3 - 8$\,keV luminosities
for the wake at core crossing of 
${\rm log}\,L_{\rm X} \sim 40.6 - 41.3$\ergs. This 
is comparable to the ${\rm log}\,L_{\rm X} = 40.9$\ergs we measure
for the X-ray trail. 
They further find that a factor two  or more enhancement in the 
surface brightness in the
wake persists out to distances $\gtrsim 60$\,kpc from the galaxy for 
times $\sim 700$\,Myr after core crossing (see Figure 14, 
Acreman \etal 2003) for galaxies with extended gas halos, making the 
$90$\,kpc long trail we observe more plausible.

In addition to the large temperature difference between the IGM and 
gas in the trail, another discrepancy between the simulations and 
our observations is that all of the two dimensional simulations that 
input spherical $\beta$-model mass distributions for the gas in the 
simulated  galaxies predict bright  X-ray enhancements at the 
center of the galaxy 
due to gas in the wake falling back and accumulating in the central 
region. While X-ray emission is observed from the central
region of NGC~6872, it is weak, far less than the orders of magnitude 
enhancement predicted by the simulations.  This may be due to
the different geometry and initial gas distribution of 
the two systems, may be an artifact of the two dimensional nature 
of the simulations (e.g., see Ruffert 1994), or may indicate the 
presence of additional physics (e.g. multiphase ISM, magnetic fields, 
nuclear feedback processes)  that have not been included  
in the simulations.  The discrepancies between our observations and 
the simulations may also be due in part to the limited spatial 
resolution of the simulations that makes it difficult 
to model well the effects of Kelvin-Helmholtz instabilities and 
turbulence on the stripping process. 

\subsection{Turbulent Viscous Stripping}
\label{sec:turbvis}

\begin{figure}[t]
\epsscale{0.5}
\begin{center}
\epsfig{file=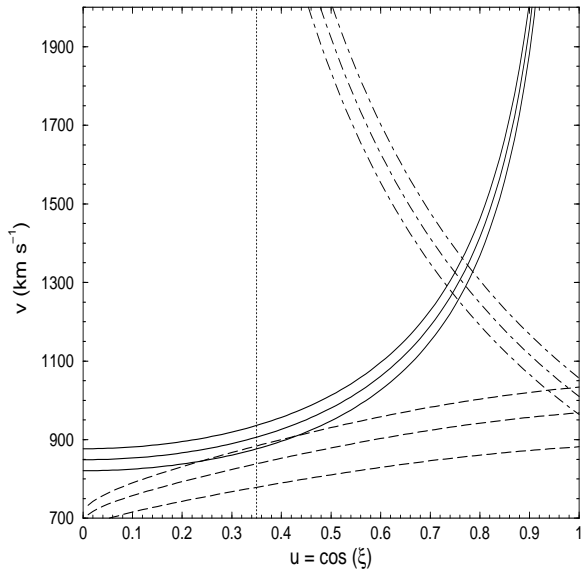, height=3in, width=3in,angle=270}
\caption{ Constraints on the velocity of NGC~6872 in the turbulent 
viscous stripping model for the trail: measured radial velocity (solid line), 
energy conservation (long-dashed line, lower bound), trail dimensions
and adiabatic expansion (dot-dashed line), mass conservation and 
turbulent viscous stripping rate (vertical dotted line, allowed 
values are to the right). Uncertainties in the constraints
are denoted by thin lines of the same kind as the constraint, and are
due to the uncertainties in the measured temperatures (long-dashed and
dot-dashed lines) and measured radial velocity (solid line).
}
\label{fig:turbvis}
\end{center}
\end{figure}

The fourth explanation for the X-ray trail  
is that it was formed by turbulent viscous stripping of cold gas 
from the spiral NGC~6872, that is thermalized and then thermally 
mixed with the ambient Pavo IGM, during the galaxy's supersonic 
motion past the group center. 
Since the mass loss rate due to turbulent viscous
stripping is largely insensitive to the orientation of the galaxy 
as it moves through the IGM, 
this process may dominate over classic ram-pressure stripping in 
large spirals (Nulsen 1982). 
Numerical simulations of this system are needed to test this 
hypothesis in detail. However, we can employ simple conservation 
arguments to determine if the scenario is qualitatively feasible.
We consider mass conservation, the turbulent-viscous stripping rate, 
energy conservation applied to the process of thermalization, and 
the constraint on the dimensions of the observed trail imposed by 
adiabatic expansion of the gas. All of these constraints depend on 
the angle $\xi$  
that the motion of NGC~6872 
makes with respect to the plane of the sky, 
either directly through the velocity of the spiral or indirectly through 
the electron densities for gas in the IGM and the trail inferred from 
observations in Section \ref{sec:dens}. 
In Figure \ref{fig:turbvis} we plot these
constraints as a function of $u = {\rm cos}(\xi)$. 

From mass conservation, the density of gas in the trail that was 
stripped from the spiral, $n_{\rm strip}$,  is given by 
$n_{\rm strip} = (\alpha -1) n_{\rm IGM}$, where $n_{\rm IGM}$ is the 
density of the undisturbed IGM and 
$\alpha = n_{\rm trail}/n_{\rm IGM}$ is the ratio of
trail to IGM gas densities. 
The rate at which mass would be 
stripped from a galaxy by turbulent viscosity,
$\dot{M}_{\rm turb}$, is given by  
\begin{equation}
 \dot{M}_{\rm turb} \sim \pi R^2 \rho_{\rm IGM} v f
\label{eq:turb}
\end{equation}
where $R$ is the radius of the gas disk ($\sim$ radius of the trail), 
$\rho_{\rm IGM}$ is the density of the ambient IGM, $v$ is the
relative velocity of NGC~6872 with respect to the IGM, and $f$
accounts for the reduction in the stripping rate when the galaxy's 
gravity becomes important (Nulsen 1982). The factor $f$ can be written
in terms of the circular velocity of the galaxy at the radius of
the gas disk, $f = {\rm min}[v^2/v^2_{\rm circ}(R),1]$. Since the
velocity of NGC~6872 through the IGM 
($v \gtrsim v_r = 849$\kms)
is larger than the estimated rotation velocity 
($v_{\rm cir} \sim 200 - 350$\kms) for the
spiral, gravity does not significantly affect the stripping rate  
and $f \sim 1$. Integrating over time in 
Equation \ref{eq:turb}, we see that turbulent viscous stripping can   
remove a gas mass from the galaxy approximately equal to the mass of
undisturbed IGM gas originally contained in the trail volume.
Turbulent-viscous stripping is efficient enough to produce the 
observed  ratio of trail to IGM gas mass provided  $\alpha \lesssim
2$.  Given the measured densities found in Section \ref{sec:dens}, 
 this restricts the angle of motion with respect to the plane of the 
sky, $\xi$, to angles less  than
$70^\circ$ ($u > 0.34$, to the right of the dotted line in 
Figure \ref{fig:turbvis}).

Energy conservation also constrains the speed of the spiral galaxy 
through the IGM in this scenario. The trail is heated to $1$\,keV 
when the kinetic
energy carried in the cold galaxy gas is thermalized by turbulent 
viscous stripping and the stripped gas mixed with the IGM. Ignoring any
initial adiabatic expansion of the gas, energy conservation implies 
\begin{equation}
\frac { \rho_{\rm strip} v^2}{2} + \frac{3 n_{\rm IGM }kT_1}{2}
\approx \frac{3 n_{\rm trail}kT_2}{2}
\label{eq:econ}
\end{equation} 
with $\rho_{\rm strip}$ the mass density of the stripped gas, 
$v$ the relative velocity of the galaxy with respect to the IGM, and 
$T_1$ and $T_2$ the temperatures of the Pavo IGM before mixing and the
X-ray trail after thermal mixing, respectively. Equation \ref{eq:econ}
can be conveniently rewritten in terms of the sound speeds $s_{\rm trail}$ 
 and $s_{\rm IGM}$ for gas in the trail and IGM, respectively, and the
ratio $\alpha$ of trail to IGM gas densities, yielding
\begin{equation}
v^2  \approx \frac{9}{5 (\alpha -1)}\bigl ( \alpha 
 s^2_{\rm trail} -s^2_{\rm IGM}\bigr )
\label{eq:esound}
\end{equation}
for gas with adiabatic index $\gamma = 5/3$. 
Since any initial work done by the gas, other dissipative forces
and adiabatic expansion were neglected, Equation \ref{eq:econ} 
(or \ref{eq:esound}) imposes a lower 
bound on NGC~6872's velocity (shown as the long-dashed lines in 
Figure \ref{fig:turbvis}, 
where the uncertainties reflect the $90\%$ CL uncertainties in 
the measured temperatures). The actual uncertainties inherent in our 
simple approximations are likely much larger.

Once the trail is formed, it is overpressured relative to the
surrounding IGM and thus adiabatically expands and cools until the 
trail gas regains pressure equilibrium with its surroundings. During 
turbulent viscous stripping the entropy of the gas may change, either
increasing, due to shocks caused by the supersonic motion of the
galaxy through the IGM, or decreasing, as a result of the thermal 
mixing of stripped gas with the ambient group medium. Numerical 
simulations are needed to determine which effect dominates. 
However, observationally
the trail appears to fade into the surrounding IGM background, so that
the final entropy of the trail is most likely close to the initial 
entropy of the IGM gas. The trail ceases to be visible because, at 
pressure equilibrium, it has expanded back nearly to the temperature 
and density of its surroundings. 
The amount of broadening in the trail is thus 
determined by the pressure ratio between the trail gas and the
surrounding IGM
\begin{equation} 
\frac{r_{\rm max}}{r_{\rm i}} \approx  
 \bigl (\frac{p_{\rm trail}}{p_{\rm IGM}}\bigr )^{1/2\gamma} =  
\bigl (\frac{\alpha T_{\rm trail}}{T_{\rm IGM}}\bigr )^{1/2\gamma} 
\end{equation}
where $r_{\rm max}$($r_{\rm i}$) are the initial and maximum  radial widths 
of the trail, $\gamma$ is the adiabatic index, $T_{\rm trail}$  
and $T_{\rm IGM}$ are the trail and IGM gas temperatures,
respectively, and $\alpha$ is the ratio of trail to IGM gas densities.
If we further assume that the 
adiabatic expansion of the trail proceeds at roughly the sound speed  
in the trail ($s_{\rm trail}$), we 
can obtain an independent estimate
of the galaxy velocity from the ratio of the physical length $l_t = l_p/u$,
given in terms of the observed projected length $l_p$ and cosine of
the angle of motion $u = {\rm cos}(\xi)$  of NGC~6872 with respect to
the plane of the sky, to its initial radial width $r_{\rm i}$
 \begin{equation}
  v \sim \frac{l_t s_{\rm trail}}{r_{\rm max}} 
  \sim \frac{l_p s_{\rm trail}}{u r_{\rm i}}\bigl (\frac{T_{\rm
IGM}}{\alpha T_{\rm trail}}\bigr )^{\frac{1}{2\gamma}}\,\,. 
\label{eq:dimsv}
\end{equation}
Equation \ref{eq:dimsv} is shown in Figure \ref{fig:turbvis} 
(dot-dashed line) for the measured projected trail length of $90$\,kpc, 
trail radial width of $33$\,kpc, and sound velocities in the trail and
IGM of $511^{+14}_{-19}$\kms and $365^{+21}_{-19}$\kms, respectively, 
corresponding to $0.98^{+0.06}_{-0.07}$\,keV trail and 
$0.50^{+0.06}_{-0.05}$\,keV IGM gas. 

Finally, the galaxy velocity must satisfy the radial velocity constraint, 
$v = v_r/sin(\xi)$, given by the solid line in Figure
\ref{fig:turbvis} for 
$v_r = 849 \pm 28$\kms (Martimbeau \& Huchra 2004).
Although these 
results need to be refined by hydrodynamical simulations, taken together 
these conditions allow us to estimate the velocity of NGC~6872 required  
to form the observed trail by turbulent-viscous stripping of the spiral. 
From Figure \ref{fig:turbvis} we find these constraints are  
satisfied if NGC~6872 moves with a speed 
$v \sim 1300$\kms at an angle $\xi \sim 40^\circ$ 
with respect to the 
plane of the sky. Once the three dimensional motion of NGC~6872 is
determined, we can complete the physical characterization of the trail.
 The ratio of trail to IGM gas densities 
is then 
$\alpha\sim 1.57$, such that $36\%$ of the trail gas has 
been stripped from the spiral with the remaining $64\%$ of the trail 
gas from the Pavo IGM.  The mean density and total gas mass in the 
trail are then $9.5 \times 10^{-4}$\cmc and $1.1 \times 10^{10}\Ms$
Given the observed projected length ($\sim 90$\,kpc) and inferred 
transverse velocity ($1000$\kms) of NGC~6872 in the plane of the sky, 
the age of the 
trail would be $\sim 90$\,Myr. We note that cooling by thermal conduction is
unimportant for the trail. The timescale for 
thermal conduction (Sarazin 1988) across the $33$\,kpc
radius of the trail at the unsuppressed Spitzer rate is 
$\sim 230$\,Myr, much  
greater than the age of the trail, and 
if tangled magnetic fields similar to those found in clusters are also
present in galaxy groups, the thermal conduction timescale may be 
$3 - 100$ times longer (Markevitch \etal 2003).

There are several key signatures of turbulent viscous stripping that can 
be tested in future observations. (1) The predicted motion of
NGC~6872 is highly supersonic 
relative to the Pavo group gas. This would produce 
a strong shock in front of the spiral galaxy that should be visible in future
X-ray observations with XMM-Newton or Chandra, in a more favorable
pointing with NGC~6872 closer to the telescope aimpoint. (2)
The metal abundance measured in the trail should be intermediate
between that in the NGC~6872 and the Pavo IGM, reflecting the mixture 
of gases in the trail. (3)
Since the trail is overpressured and adiabatically expands, a simple 
prediction of this scenario is that, as the distance from the
spiral increases, the trail should broaden and become cooler, less
dense and dimmer. In Figure \ref{fig:rprof} we see evidence that this
may be the case. The difference between the surface brightness
observed in the trail (northwest) region and IGM (southern) region 
decreases with increasing distance from the spiral, i.e. as one moves
closer to the dominant elliptical NGC~6876. (4) The temperature
profile along the trail as a function of distance from NGC~6872 is 
distinctive. Since the X-ray trail is heated as the kinetic energy
carried in the stripped, cold galaxy gas is thermalized and mixed with
the IGM, this model predicts a cooler temperature near NGC~6872 where
the thermalization is incomplete, in contrast to the hotter
temperatures expected for a Bondi-Hoyle wake. The temperature is
predicted to rise initially with distance as thermalization and mixing
is complete, and then decrease with the adiabatic expansion of the gas.
(5) The observation of filamentary features in the trail, as suggested
in Figure \ref{fig:pavo}, would also 
favor turbulent-viscous stripping, since the parent cold gas
clouds need not be uniformly distributed in the spiral. Furthermore, 
the interaction of the companion galaxy IC~4970 with NGC~6872 may also 
displace gas within the spiral, making it easier to be stripped.

While physically possible, the velocity $1300$\kms (Mach $\sim 3.6$) 
of NGC~6872 required to satisfy both the radial velocity constraint
and the geometrical constraint in
Equation \ref{eq:dimsv}, imposed by the dimensions of the trail, is 
unusually high for motion through a cool group. This is a result of 
the competing factors imposed by the 
long observed projected length of the trail and the rapid adiabatic 
expansion of the overpressured trail gas, at the speed of 
sound $s_{\rm trail}$, after thermal mixing. However, as discussed in 
Section \ref{sec:wake}, the more likely physical situation is that 
Bondi-Hoyle  gravitational focusing is also significant and acts to 
inhibit the adiabatic expansion of the trail gas. Thus the effective
expansion velocity of the trail in Equation \ref{eq:dimsv} 
decreases so that the time over which the 
trail is visible (trail age) increases, shifting the dot-dashed curve
in Figure \ref{fig:turbvis} to lower galaxy velocities.

\section{Conclusions}
\label{sec:conclude}

The observation of X-ray wakes and trails behind galaxies in groups
and clusters opens exciting new opportunities to constrain the  
dynamical motion  of galaxies in these systems and to 
probe the physical processes that govern their interaction with the 
surrounding environment. In this paper we have presented the results
of a $32.2$\,ks XMM-Newton observation of an X-ray trail spanning the 
full $8'.7 \times 4'$ projected area between the 
large spiral galaxy NGC~6872 and the dominant elliptical galaxy 
NGC~6876 in the Pavo group, the first observation of an X-ray 
trail produced by a spiral galaxy in a poor galaxy group.

In summary we found:
\begin{enumerate}

\item{The $0.5-2$\,keV surface brightness profile for the dominant 
elliptical NGC~6876 is well described by a $\beta$-model with 
core radius $r_c = 5$\,kpc and index $\beta = 0.65$ within $10$\,kpc. 
Beyond $10$\,kpc the surface brightness becomes highly asymmetric with
the profile to the south (away from the spiral NGC~6872) well
described by the addition of a second Pavo IGM component with 
$r_c = 50$\,kpc, $\beta=0.3$; while to the northwest (in the direction
of NGC~6872) the surface brightness beyond $\sim 20$\,kpc is constant 
or slowly rising, and by $r \sim 60$\,kpc is a factor $\gtrsim 2$
higher than in the Pavo IGM to the south.
}

\item{Using a single temperature APEC model with Galactic absorption,
we find a temperature and abundance 
for the undisturbed Pavo IGM of $0.50^{+0.06}_{-0.05}$\,keV 
and $0.05^{+0.03}_{-0.02}\,\Zs$, 
and a group $0.5-2$\,keV luminosity (including NGC~6876)
within a radius of $120$\,kpc of $6.3 \times 10^{41}$\ergs.
}

\item{ The spectrum of gas in the (northwest) X-ray trail region is
well fit by a two temperature model with one component fixed by the 
Pavo IGM background and the other varied to determine the
properties of the X-ray gas in the trail. We find a temperature
$0.98^{+0.08}_{-0.07}$\,keV and abundance $0.2 \pm 0.1\,\Zs$ 
for gas in the trail. The $0.5-2$\,keV ($2-10$\,keV) luminosity of 
the X-ray trail component from the $4'.35 \times 5'.9$ rectangular 
 extraction region is $6.6 \times 10^{40}$\ergs 
($9 \times 10^{39}$\ergs). 
 }

\item{For the dominant elliptical galaxy, NGC~6876, the spectrum 
for a $99''.7$ circular region (excluding NGC~6877) is best fit by 
a two temperature APEC model with abundance 
$0.95 \pm 0.3\,\Zs$ and temperatures
$0.75\pm 0.02$  and $1.6 \pm 0.1$\,keV for fixed Galactic 
absorption. We find $0.5-2$ ($2-10$) luminosities of 
$2.4 \times 10^{41}$\ergs ($3.7 \times 10^{40}$\ergs), a central 
electron density of $\sim 0.02$\cmc, and hot gas mass 
$\sim 3.4 \times 10^9\Ms$. 
}

\item{The spectrum for the  $58''.6$ central region of the spiral
NGC~6872 is best fit using an APEC plus power law model with 
temperature $kT =0.65 \pm 0.06$\,keV and photon index 
$\Gamma = 2.0^{+0.3}_{-0.4}$ ($\Gamma = 1.3^{+0.3}_{-0.2}$) 
assuming fixed extremes for the abundance, $A=1\,\Zs$ 
($A=0.2\,\Zs$), respectively. The $0.5-2$\,keV luminosity from this 
region is $4.9 \times 10^{40}$\ergs ($4.6 \times 10^{40}$\ergs) for 
these models. The $2-10$\,kev luminosity, $2.9 \times 10^{40}$\ergs 
($4.0 \times 10^{40}$\ergs) is too high to 
be produced by LMXBs alone. If attributed to HMXBs, the $2-10$\,keV
luminosity predicts a star formation rate in the central $58''.6$ 
of the spiral NGC~6872 of $\sim 4.3\Ms\,{\rm yr}^{-1}$ 
($\sim 6\Ms\,{\rm yr}^{-1}$).
}

\item{The measured subsolar abundance for gas in the X-ray trail
is similar to that of the undisturbed Pavo IGM. However, better 
spectra  are needed to measure abundances in the spiral 
NGC~6872, as well as in the IGM and trail, to determine the origin
of gas in the trail.
}

\item{The trajectory for NGC~6872, 
inferred by the presence of the trail, suggests that tidal interactions
with NGC~6876, as well as those from its interacting companion
IC~4970, may have contributed to 
the tidal distortion of the galaxy's spiral arms and HI gas
distribution. These interactions also should be included in dynamical 
models of the spiral galaxy's evolution.
}
 
\item{A possible explanation for the X-ray trail 
is that it is thermally mixed Pavo IGM gas ($64\%$) and 
ISM gas ($36\%$) that has been stripped from NGC~6872 by 
turbulent viscosity as the spiral moves supersonically
($v \sim 1300$\kms) through the Pavo IGM 
at an angle $\xi \sim 40^\circ$ with respect to the plane of the sky, 
passing  the Pavo group center (the elliptical NGC~6876) $\sim 130$\,Myr
ago. The dimming of the trail as the gas adiabatically expands 
can explain the $\sim 100$\,kpc projected length of the trail. Better
simulations that include the effects of turbulence on stripping 
processes are needed to model the complex dynamics of this system.
}

\item{ The mean electron density and total mass in the trail inferred 
from the data are dependent on the projection geometry.
Assuming uniform filling of the cylindrical geometry and an 
angle $\xi \sim 40^\circ$, from the turbulent-viscous stripping model, 
for the motion of the spiral with respect to the plane of the sky, 
we find a mean electron density for gas in the X-ray trail of $9.5 \times
10^{-4}$\cmc and total hot gas mass in the trail
of $1.1 \times 10^{10}\Ms$. Nonuniform filling, as suggested by the 
X-ray images, would increase (reduce) the density (mass) by factors
$\eta^{-1/2}$ ($\eta^{1/2}$), respectively, where $\eta \leq 1$ is the 
filling factor.
}

\item{ Gravitational focusing of IGM gas into a Bondi-Hoyle wake 
due to the highly supersonic motion of NGC~6872 through the Pavo IGM 
may also be significant. Better X-ray observations are needed to 
measure the metal
abundance and temperature profile near the trail head to distinguish 
between gravitational focusing and turbulent-viscous stripping as the 
dominant trail formation mechanism and to find the bow shock predicted
by both models.  Optical measurements are 
needed to measure the rotation curve in the spiral NGC~6872. Better 
numerical simulations are needed to constrain the properties of  
NGC~6872's dark matter halo and thus set the scale for 
gravitational focusing in this system.}

\end{enumerate}


\acknowledgements

 This work has been supported in part by NASA contract G03-4176A,
 and the Smithsonian Institution. MEM and LS also
  acknowledge support from the Radcliffe Institute for Advanced Study
  at Harvard University. 
 This work has made use of the NASA/IPAC Extragalactic Database (NED)
 which is operated by the Jet Propulsion Laboratory, California
 Institute of Technology,  under contract with the National
 Aeronautics and Space Administration. We wish to thank 
 John Huchra and Michael McCullough for useful discussions, and 
 Ralph Kraft for help with the data.


\appendix
\section{Gas flow through an NFW potential}
The model is of an ``empty'' (gas free) galaxy moving through an
otherwise uniform gas.  Cooling can be ignored, since the time for gas
to pass through the wake is significantly shorter than its cooling
time.  The effect of gravity is treated as a perturbation, using
linearized flow equations.  This gives an accurate result when the
fractional density perturbation is small, which is a marginal
approximation in practice.

The full flow equations are
\begin{eqnarray}
{d\rho\over dt} + \rho \nabla \cdot {\bf v} = 0, \label{eqn:mass} \\ 
\rho {d{\bf v} \over dt} = -\nabla p + \rho {\bf g} \label{eqn:mom}
\end{eqnarray}
and
\begin{equation}
S = {\rm constant}, \label{eqn:entrop}
\end{equation}
where $\rho$, $p$, $S$ and $\bf v$ are the density, pressure,
specific entropy and velocity, respectively, of the gas.  The
acceleration due to gravity is $\bf g$ and the lagrangian derivative
is $d/dt = \partial/\partial t + {\bf v} \cdot \nabla$.  In a frame
moving with the galaxy, the flow is steady ($\partial/\partial t =
0$).  Gas properties at large distances from the galaxy are unform and
we denote them by subscript `0'.  Treating the effect of gravity as a 
perturbation, we write $\rho = \rho_{g0} + \delta \rho$, $p = p_0 +
\delta p$ and ${\bf v} = {\bf v}_0 + \delta {\bf v}$, and linearize the
flow equations.  Equation (\ref{eqn:mass}) then gives 
\begin{equation}
{\bf v}_0 \cdot \nabla f + \nabla \cdot \delta {\bf v} = 0,
\label{eqn:lmass} 
\end{equation}
where $f = \delta\rho / \rho_{g0}$ is the fractional density perturbation.
Equation (\ref{eqn:entrop}) implies that $\delta p = s_0^2 \, \delta
\rho$, where $s_0$ is the speed of sound in the unperturbed gas, so
that the linearized form of equation (\ref{eqn:mom}) can be written as
\begin{equation}
{\bf v}_0 \cdot \nabla \delta {\bf v} = - s_0^2 \nabla f + {\bf
  g}. \label{eqn:lmom} 
\end{equation}
Eliminating $\delta {\bf v}$ between (\ref{eqn:lmass}) and
(\ref{eqn:lmom}) gives
\begin{equation}
({\bf v}_0 \cdot \nabla)^2 f - s_0^2 \nabla^2 f = - \nabla \cdot {\bf
    g} = 4 \pi G \rho_\ast, \label{eqn:key}
\end{equation}
where $\rho_\ast$ is the density of gravitating matter in the galaxy
(the gas is assumed to have negligible gravity).

Equation (\ref{eqn:key}) for the density perturbation is linear, with
the gravitating mass density as its source terrm.  Its solution is
conveniently expressed in terms of a Greens function, $q({\bf r})$,
a solution of 
\begin{equation}
({\bf v}_0 \cdot \nabla)^2 q - s_0^2 \nabla^2 q = 4 \pi G M \delta({\bf
    r}), \label{eqn:rs} 
\end{equation}
where $\delta ({\bf r})$ is the Dirac delta function.  The solution of
(\ref{eqn:key}) is then
\begin{equation}
f({\bf r}) = {1 \over M} \int q({\bf r} - {\bf r}') \rho_\ast({\bf
  r}') \, d^3{\bf r}'. \label{eqn:gfsol}
\end{equation}
Equation (\ref{eqn:rs}) is the linearized equation for the density
perturbation due to a point mass, $M$, moving through a uniform,
adiabatic gas (Ruderman \& Spiegel 1971).  For a point mass moving
supersonically in the $+z$ direction, with Mach number $m = v_0/s_0$,
its solution expressed  in terms of the radius, $r$, and polar angle,
$\theta$, is 
\begin{displaymath}
q({\bf r}) = {2 G M \over s_0^2 r \sqrt{1 - m^2 \sin^2 \theta}},
\qquad {\rm for\ } \cos\theta < - \sqrt{1 - 1 / m^2}
\end{displaymath}
and $q({\bf r}) = 0$ otherwise (outside the Mach cone).  The solution
for subsonic motion is formally half of this, but non-zero everywhere,
so that the density perturbation is the same in front of and behind
the perturber.  We only consider the supersonic case below.

Using these results, the fractional density perturbation at a point on
the $z$ axis, at ${\bf r} = (0,0,z)$, due to a galaxy moving at Mach
$m$ in the $+z$ direction is given by
\begin{displaymath}
f(z) = \int {2 G \rho_\ast({\bf r}') \, d^3 {\bf r}' \over s_0^2
  \sqrt{ (z - z')^2 - (m^2 - 1) (x'^2 + y'^2) }},
\end{displaymath}
integrated over the interior of the inverted Mach cone of $(0,0,z)$
(i.e.\ the region defined by $z' - \sqrt{(m^2 - 1) (x'^2 + y'^2)} >
z$).  If the distribution of gravitating matter is spherically
symmetric, with $\rho_\ast({\bf r}') = \rho_\ast(r')$, then this can
be integrated over the angular coordinates, giving
\begin{equation}
f(z) = {4\pi G\over s_0^2} \int_0^\infty g(m, z/r') \rho_\ast(r') r'
\, dr', \label{eqn:dpsph}
\end{equation}
where
\begin{displaymath}
g(m, t) = 
\begin{cases}
{1\over m} \ln \left| m - t \over m + t \right|, &t < -1, \cr
{1\over 2m} \ln {(m + 1) (m - t) \over (m - 1) (m + t)}, &-1 < t < 1,
\cr
0, & t>1. \cr
\end{cases}
\end{displaymath}
Taking $\rho_\ast(r) = \rho_0 / [(r/a)(1 + r/a)^2]$ for $r/a < c$,
appropriate to an NFW potential with inner radius $a$ and  
concentration parameter $c$, equation (\ref{eqn:dpsph}) gives
\begin{displaymath}
f(z) = {4\pi G \rho_0 a^2 \over s_0^2}
\begin{cases}
{1 \over m (1 + c)} \ln \left|mc + z/a \over mc - z/a \right|
  + {1 \over z/a + m} \ln \left| m c - z/a \over (1 + c) z/a \right| 
\cr\qquad\qquad\qquad\qquad\qquad
  + {1 \over z/a - m} \ln \left| m c + z/a \over (1 + c) z/a \right|, 
& z/a < -c \cr
{1 \over 2m (1 + c)} \ln {(m-1)(mc + z/a) \over (m+1)(mc - z/a)}
  + {1\over 2(z/a+m)}\ln\left|(m+1)(mc-z/a)\over(1+c)(1-z/a)z/a\right|
\cr\qquad\qquad\qquad\qquad\qquad
+ {1\over 2(z/a-m)}\ln\left|(m-1)(mc+z/a)\over(1+c)(1-z/a)z/a\right|, 
& -c < z/a < 0 \cr
{1 \over 2m (1 + c)} \ln {(m-1)(mc + z/a) \over (m+1)(mc - z/a)}
+ {1\over 2(z/a + m)} \ln {(mc - z/a)(1 + z/a) \over (m-1)(1+c) z/a} 
\cr\qquad\qquad\qquad\qquad\qquad
+ {1\over 2(z/a - m)} \ln {(mc + z/a)(1 + z/a) \over (m+1)(1+c) z/a},
& 0 < z/a < c \cr
0, & z/a > c.\cr
\end{cases}
\end{displaymath}
 

\end{document}